\documentclass[12pt]{iopart}
\usepackage{iopams}
\gdef\ffrac#1#2{\textstyle{#1\over#2}\displaystyle}
\gdef\zb{{\bar z}}
\gdef\hb{{\bar h}}
\gdef\O{{\Phi}}
\gdef\be{\begin{equation}}
\gdef\ee{\end{equation}}
\eqnobysec

\makeatletter
\renewcommand\tableofcontents{%
    \section*{\contentsname}   
   \@starttoc{toc}%
   }
\makeatother

\begin{document}
\title[LogCFTs as limits of ordinary CFTs]{Logarithmic conformal field theories as limits of ordinary CFTs and some physical applications}
\author{John Cardy$^{1,2}$}
\address{$^1$Rudolf
Peierls Centre for Theoretical Physics, 1 Keble Road, Oxford OX1
3NP, UK}
\address{$^2$All Souls College, Oxford}

\begin{abstract}
We describe an approach to logarithmic conformal field theories as limits of sequences of ordinary conformal field theories with varying central charge $c$. Logarithmic behaviour arises from degeneracies in the spectrum of scaling dimensions at certain values of $c$. The theories we consider are all invariant under some internal symmetry group, and logarithmic behaviour occurs when the decomposition of the physical observables into irreducible operators becomes singular. Examples considered are quenched random magnets using the replica formalism, self-avoiding walks as the $n\to0$ limit of the $O(n)$ model, and percolation as the limit $Q\to1$ of the Potts model. In these cases we identify logarithmic operators  and pay particular attention to how the $c\to0$ paradox is resolved and how the $b$-parameter is evaluated. We also show how this approach gives information on logarithmic behaviour in the extended Ising model, uniform spanning trees and the $O(-2)$ model. Most of our results apply to general dimensionality. 

We also consider massive logarithmic theories and, in two dimensions, derive sum rules for the effective central charge and the $b$-parameter. 

\end{abstract}
\maketitle

\tableofcontents

\section{Introduction}\label{sec1}
Multiplicative logarithmic corrections to power law behaviour at or near critical points are normally associated with the presence of operators in the effective hamiltonian which are marginally irrelevant under the renormalisation group (RG). Such operators may give rise, for example, to behaviour of correlation functions at the critical point of the form \cite{Weg}
\be
\langle\Phi(r)\Phi(0)\rangle\sim\frac a{r^{2x}}\,(1+g\log r)^{x'}\,,
\ee
where $a$ is an amplitude, $x$ is the scaling dimension of $\Phi$, $x'$ is some universal ratio of operator product expansion (OPE) coefficients, and $g$ measures the strength of the marginally irrelevant operator. 
However, the amplitudes of such logarithms vanish at the RG fixed point when $g=0$. 

RG fixed points in rotationally invariant systems with short-range interactions correspond to conformal field theories (CFTs). It was pointed out some time ago \cite{Sal,Gur} that the structure of general CFTs allows the presence of multiplicative logarithms in correlation functions even at an RG fixed point. Such CFTs are called logarithmic (logCFTs). Most work on the subject since then has focussed on elucidating properties of logCFTs directly by attempting to generalise notions of ordinary CFTs. However this has proved difficult because they are less constrained than unitary rational CFTs. Moreover, while it is often possible to identify unitary rational CFTs as scaling limits of local observables of physical systems, the corresponding arguments for logCFTs are lacking. Therefore, while it is known that important physical systems (for example the quantum Hall plateau transition \cite{QH}) should be described at their critical point by a logCFT, it has proven very difficult to identify which logCFT this should be.

In this article we adopt a different, complementary approach, considering logCFTs as limits of ordinary, non-logarithmic (albeit irrational) CFTs, whose physical interpretation is already well understood, as a parameter is taken to a particular value. In this way we are able to derive properties of the corresponding logCFTs, and to understand exactly where the logarithms should appear in physical observables. While our approach is quite general, we illustrate it with several examples of physical interest, including quenched random magnets, self-avoiding walks, and percolation. This article is a considerably expanded and updated version of Refs.~\cite{Cardy1,Cardy2}. Some of the material appeared in Ref.~\cite{Kogan}.

To see how logarithmic factors in correlation functions might arise when taking a suitable limit of a conventional CFT, consider the general form of a two-point function in such a theory. A given local observable $\Phi$ is expanded in scaling operators $\phi_i$, each of which has a unique scaling dimension $x_i$. Its two-point function in ${\mathbb R}^d$ therefore has the form
\begin{equation}\label{e1}
\langle\Phi(r)\Phi(0)\rangle\sim\sum_{ij}\frac{a_{ij}}{r^{x_i+x_j}}\,.
\end{equation}
Conformal invariance implies that $a_{ij}=0$ when $x_i\not=x_j$ (and the $\phi_i$ are quasi-primary) \cite{orthog}.
Moreover, within each subspace with $x_i=x_j$, and for a \em unitary \em CFT, the matrix $a_{ij}$ is positive definite. Unitarity implies reflection positivity of correlation functions, which is expected to hold for all local operators $\Phi$ in models with local positive Boltzmann weights \cite{FQS}. This rules out logarithmic behaviour in such cases.

However, not all correlation functions in interesting theories necessarily satisfy the condition of reflection positivity. Examples, to be discussed in detail later, are models such as the $O(n)$ model and the $Q$-state Potts model, which can be written in terms of non-intersecting random loops, in terms of which both the observables and the Boltzmann weights are non-local. Another set of examples arises in disordered systems, where the use of the replica trick involves taking $n$ coupled copies of the original system, then taking the limit $n\to0$. In all these cases the parameter $n$ or $Q$ may take a continuous set of values and the values of the scaling dimensions will in general depend continuously on this parameter. 

Under these circumstances it is easy to see how logarithmic behaviour might arise. For suppose that, as $n$ (or $Q$) tends to some particular value $n_c$, a pair of scaling dimensions $x$ and $\tilde x$ collide, that is $x-\tilde x\to0$, and at the same time the corresponding amplitudes
$a\sim-\tilde a\to\infty$ in such a way that $a(x-\tilde x)$ remains finite. Then the leading terms in (\ref{e1}) cancel, leaving behind a term proportional to $r^{-2x}\log r$. (If more than two scaling dimensions collide at the same value of $n$, higher powers of $\log r$ may arise.) More explicitly, suppose that
\begin{eqnarray}
\langle\phi(r)\phi(0)\rangle=\frac{A(n)}{n-n_c}\,r^{-2x(n)}\,,\\
\langle\tilde\phi(r)\tilde\phi(0)\rangle=-\frac{\widetilde A(n)}{n-n_c}\,r^{-2\tilde x(n)}\,,\\
\langle\phi(r)\tilde\phi(0)\rangle=0\,,
\end{eqnarray}
where $A(n)$ and $\widetilde A(n)$ have the same finite limits as $n\to n_c$, and $x(n),\tilde x(n)$ are differentiable at $n_c$. The last equation follows from conformal invariance if $x(n)\not=\tilde x(n)$.
Defining 
\be\label{CDdef}
D\equiv\phi-\tilde\phi\,,\qquad C\equiv\big(x(n)-\tilde x(n)\big)\phi\,, 
\ee
then, as $n\to n_c$,
\begin{eqnarray}
\langle D(r)D(0)\rangle=\frac{A(n)}{n-n_c}\frac1{r^{2x(n)}}-\frac{\widetilde A(n)}{n-n_c}\frac1{r^{2\tilde x(n)}}
\to-\frac{2\alpha\big(\log r+O(1)\big)}{r^{2x(n_c)}}\,,\label{e17}\\
\langle C(r)D(0)\rangle=\frac{A(n)(x(n)-\tilde x(n))}{n-n_c}\frac1{r^{2x(n)}}
\to\frac{\alpha}{r^{2x(n_c)}}\,,\\
\langle C(r)C(0)\rangle=\frac{A(n)(x(n)-\tilde x(n))^2}{n-n_c}\frac1{r^{2x(n)}}
\to0\,,\label{e19}
\end{eqnarray}
where
$\alpha=\big(x'(n_c)-\tilde x'(n_c)\big)A(n_c)$.

$(C,D)$ form a \em logarithmic pair \em\cite{Gur}. In radial quantisation, if $\cal D$ is the generator of dilatations, for $n\not=n_c$
we have ${\cal D}|\phi\rangle=x|\phi\rangle$ and ${\cal D}|\tilde\phi\rangle=\tilde x|\tilde\phi\rangle$, so that 
\begin{eqnarray}
{\cal D}|C\rangle&=x|C\rangle\,,\\
{\cal D}|D\rangle&=x|\phi\rangle-\tilde x|\tilde\phi\rangle=|C\rangle+\tilde x(n)|D\rangle\,,\\
&\to|C\rangle+x|D\rangle\,.
\end{eqnarray}
Thus $(C,D)$ span a Jordan cell of $\cal D$:
\be
{\cal D}\,
\left(\begin{array}{c}C \\D \end{array}\right)=
\left(\begin{array}{cc}x&0  \\1& x \end{array}\right)\left(\begin{array}{c}C \\D \end{array}\right)\,.\label{e113}
\ee
We see this is formed by the \em collision \em of the operators $\phi$ and $\tilde\phi$ as $n\to n_c$. Higher rank Jordan cells can arise from the collision of more than two operators. 

It may seem somewhat artificial to suppose that the amplitudes $a(n)$ and $\tilde a(n)$ should become singular in this particular way. However we shall argue that this occurs naturally when the theory possesses an internal symmetry.

\subsection{The role of internal symmetries}

The examples we shall consider in this review have the further property that the effective hamiltonian is invariant under some group ${\cal G}_n$ of global internal symmetries. For the $O(n)$ model this is the orthogonal group, for the Potts model and replica models it is the permutation group $S_n$. In general $n$ will index either the number of generators of the group or its order and thus should initially be a positive integer. Since the elements of ${\cal G}_n$ commute with the generator $\cal D$ of scale transformations, for these values of $n$ the scaling operators $\phi_{j;\alpha}$ should transform according to irreducible representations of ${\cal G}_n$. We shall be interested in representations carrying a finite number of indices, denoted by $\alpha$, that are classified by Young tableaux with a fixed number of rows. These have the property that the correlation functions of operators transforming according to these representations can be continued to non-integer values of $n$, even though the group elements themselves have no direct meaning.  Scaling operators correspond to well-defined eigenvalues $x_j(n)$ of $\cal D$, so that their 2-point functions take the form
\be\label{114}
\langle\phi_{i;\alpha}(r)\phi_{j;\beta}(0)\rangle=\delta_{ij}A_{j;\alpha\beta}(n)/r^{2x_j(n)}\,,
\ee
where the form of $A_{j;\alpha\beta}(n)$ is fixed by group theory.

However, it may well be that the physical operators $\Phi$ are \em not \em irreducible under ${\cal G}_n$, but rather may be written as a linear combination
\be\label{115}
\Phi(r)=\sum_{j,\alpha}f_{j;\alpha}^\Phi(n)\phi_{j;\alpha}(r)\,,
\ee
where the coefficients are, up to an overall multiplicative factor, again completely determined by group theory.
Hence
\begin{equation}\label{PhiPhi}
\langle\Phi(r)\Phi(0)\rangle=\sum_j\sum_{\alpha,\beta}\frac{f_{j;\alpha}^\Phi(n)f_{j;\beta}^\Phi(n)A_{j;\alpha\beta}(n)}
{r^{2x_j(n)}}\,.
\end{equation}

For a unitary CFT $A_{j;\alpha\beta}(n)$ is positive definite, but this no longer holds true for general values of $n$.
The logarithmic behaviour we consider here happens at values of $n$ for which $A_{j;\alpha\beta}(n)$ (and also the $f_{j;\alpha}^\Phi(n)$) are singular. If there is a physical requirement that the correlator on the left hand side of (\ref{PhiPhi})
be finite at this value of $n$, then these singularities must cancel between the various terms on the right hand side.
As explained above, this can then lead to logarithmic terms.

A simple example, to be discussed in more detail in Sec.~\ref{sec2}, is as follows. Suppose the group ${\cal G}_n$ 
is $S_n$, and we consider a set of operators $\Phi_a$ with $a\in(1,\ldots,n)$ on which the elements $\pi$ of the group act in the obvious way: $\Phi_a\to\Phi_{\pi(a)}$. The irreducible operators are
\be
\Phi\equiv\sum_{a=1}^n\Phi_a\,,\qquad
\widetilde\Phi_a\equiv\Phi_a-(1/n)\sum_{a=1}^n\Phi_a\quad\mbox{with} \sum_{a=1}^n\widetilde\Phi_a=0\,,
\ee
transforming according to the singlet and $(n-1)$-dimensional representation respectively.
In general $\Phi$ and $\widetilde\Phi_a$ will have different scaling dimensions $x(n)$ and $\tilde x(n)$, so they are orthogonal by conformal invariance. On the other hand, since $\sum_1^n\widetilde\Phi_a=0$, the 2-point functions of these operators have the forms dictated by group theory
\begin{equation}\label{32}
\langle\widetilde\Phi_a(r)\widetilde\Phi_b(0)\rangle=\frac{\widetilde A(n)\,(\delta_{ab}-1/n)}{r^{2\tilde x(n)}}\,,\qquad
\langle\Phi(r)\Phi(0)\rangle=\frac{A(n)\,n}{r^{2x(n)}}\,,
\end{equation}
thus defining the amplitudes $\widetilde A(n)$ and $A(n)$. We expect these to be finite as $n\to0$ by studying the behaviour in some reference theory, eg when fields $\Phi_a$ are independent so that $\langle\Phi_a\Phi_b\rangle\propto\delta_{ab}$. 

In this example both the coefficients $f$ and $A$ as defined in (\ref{114},\ref{115}) are singular at $n=0$. 
The 2-point functions of the physical fields $\Phi_a$  are however
\begin{eqnarray}
\langle\Phi_a(r)\Phi_a(0)\rangle=\frac{\widetilde A(n)\,(1-1/n)}{r^{2\tilde x(n)}}+\frac1{n}\frac{A(n)}{r^{2x(n)}}\,,\label{p1}\\
\langle\Phi_a(r)\Phi_b(0)\rangle=-\frac1n\frac{\widetilde A(n)}{r^{2\tilde x(n)}}+\frac1{n}\frac{A(n)}{r^{2x(n)}}
\,,\quad(a\not=b)\,.\label{p2}
\end{eqnarray}
There is an obvious potential problem if we want to consider the limit $n\to0$, which, as we shall discuss later, is the physical limit in disordered systems. One way around this would be to suppose that both $\widetilde A$ and $A$ are $O(n)$.  However, this would imply that in the limit $\langle\Phi_a\Phi_a\rangle-\langle\Phi_a\Phi_b\rangle$
vanishes, and, as we shall argue, this cannot be true for an interacting system. 

A less severe option is to suppose that  $\widetilde A\sim A=O(1)$, so that the leading terms on the right-hand side of (\ref{p1},\ref{p2}) can cancel. As explained above, this will then lead to logarithmic behaviour if $\tilde x'(0)\not=x'(0)$. In Sec.~\ref{sec2} we shall argue that this is indeed the case for certain operators in disordered systems.

In this example one may identify the logarithmic pair $(C,D)$ normalised as in (\ref{e17}-\ref{e19}):
\be
C=\frac{x(n)-\tilde x(n)}n\,\Phi\,;\qquad D=\widetilde \Phi_a+\frac1n\Phi=\Phi_a\,.
\ee
We see that the logarithmic field $D$ should actually carry an index, $D_a$, and this then transforms reducibly under $S_n$. In particular (\ref{e17}) should read
\be
\langle D_a(r)D_b(0)\rangle\sim-\frac{2\alpha(\log r+O(1))}{r^{2x(n_c)}}+\frac{A(n_c)\delta_{ab}}{r^{2x(n_c)}}\,.
\ee
We see that although the logarithmic terms are blind to the group structure, there is still a remnant of this in the non-logarithmic correction. Moreover the ratio of the amplitude of the the non-logarithmic correction to that of the logarithmic term is universal $\propto (x'(0)-\tilde x'(0))^{-1}$ and cannot be gauged away.

\subsection{The `$c\to0$ paradox'}
We have seen that logarithms may occur in two-point functions. They may arise similarly in higher-point functions, if coefficients in the short-distance operator product expansion (OPE) become singular. It was pointed out in 
Ref.~\cite{Gur2} that this occurs generically for CFTs with central charge $c=0$. Since $c$ is related to the response of the partition function $Z$ to a scale transformation, any model with $Z=1$ necessarily corresponds to such a CFT. This includes several important examples, such as the $O(n)$ model as $n\to0$, corresponding to self-avoiding walks, the $Q$-state Potts model as $Q\to1$, corresponding to percolation, and quenched disordered systems, described either by the $n\to0$ replica trick or using supersymmetry. 

In a generic CFT with $c\not=0$, the OPE of a scaling operator $\phi$ with itself takes the form
\begin{equation}\label{TOPE}
\phi(r)\cdot\phi(0)=\frac{a_\phi}{r^{2x_\phi}}\left(1+B\frac{x_\phi}{c}r^d\,T(0)+\cdots\right)\,.
\end{equation}
Here $a_\phi$ is the normalisation of the 2-point function, $B$ is a calculable constant, and the symbol $T$ represents the stress tensor $T^{\mu\nu}$. This is valid in any number of dimensions $d$, but in general there is a complicated coordinate index structure \cite{JCTT} which has been suppressed. In this context, the central charge $c$ is defined\footnote{This is valid for all $d$. However, for $d>2$, $c$ defined in this manner is not directly related to the coefficient $a$ of the Euler term in a curved background, which for even $d$  satisfies a generalisation of the $c$-theorem \cite{Kom}.} in terms of the 2-point function of $T$
\be
\langle T(r)T(0)\rangle=\frac c{r^{2d}}\times\mbox{index structure}\,.
\ee
The coefficient of $T$ in (\ref{TOPE}) then follows by taking the correlator of both sides with $T(r')$ and using the above equation and the conformal Ward identity which implies that $\langle T(r')\phi(r)\phi(0)\rangle\propto x_\phi$. 

If the OPE (\ref{TOPE}) is inserted into a 4-point function, it takes the form
\be
\langle\phi(r_1)\phi(r_2)\phi(r_3)\phi(r_4)\rangle\sim\frac{a_\phi^2}{r_{12}^{2x_\phi}r_{34}^{2x_\phi}}
\left(1+\cdots+B^2\frac{x_\phi^2}{c}\eta^d+\cdots\right)\,,
\ee
in the limit when $r_{12}\sim r_{34}\ll r_{13}\sim r_{24}$. Here $\eta=r_{12}r_{34}/r_{13}r_{24}$ is the cross-ratio.

There is an obvious potential problem if $c=0$. In the examples considered here, this may be studied by taking the limit $n\to0$ or $Q\to1$ so that $c\to0$. The paradox may then be resolved in at least three possible ways:

\begin{itemize}
\item[a)] the physical normalisation $a_\phi$ of the field $\phi$ vanishes as $c\to0$: this happens for the field $\Phi$ in (\ref{32}), for the order parameter in percolation (see Sec.~\ref{sec4}), for bulk Kac operators in $d=2$ (see Sec.~\ref{sec13}), and for the operator $C$ in any logarithmic pair $(C,D)$ (see (\ref{e19}));
\item[b)] the scaling dimension $x_\phi\to0$: this happens in $d=2$ for the $\phi_{1,2}$ Kac operator, whose 4-point function gives the crossing formula in percolation \cite{JCcrossing};
\item[c)] there is an omitted term in (\ref{TOPE}) of the form 
\be
\widetilde B\frac{x_\phi}cr^{d+\delta}\widetilde T(0)\,,
\ee
with same spatial index structure as the term involving $T$, where, for $c\not=0$, $\widetilde T$ is a traceless symmetric tensor (in space) with scaling dimension $d+\delta$, which collides with $T$ as $\delta=O(c)\to0$, in such a way that ${\widetilde B}^2\to B^2$ and $\langle\widetilde T(r)\widetilde T(0)\rangle\to-c/r^{2d}$. This will then give rise to logarithmic terms $\propto \eta^d\log\eta$ in the  4-point function. As we shall argue, this is what happens for the operators $\tilde\Phi_a$ in the replica approach (see Sec.~\ref{sec2}), and more generically in logCFTs.
\end{itemize}

In case c), if  we additionally choose the normalisation of $\widetilde T$ so that $\widetilde B\to-B$ in the limit, 
and define
\be
t\equiv(2/\delta)(\widetilde T-T)\,,
\ee
then the OPE (\ref{TOPE}) becomes, in the limit,
\be
\fl\phi(r)\cdot\phi(0)=\frac{a_\phi}{r^{2x_\phi}}\left(1+B\frac{x_\phi}{b}r^d\,t(0)+B\frac{x_\phi}{b}r^d\log r\,T(0)+B'x_\phi r^dT(0)+
\cdots\right)\,,
\ee
where
\be
b\equiv-\lim_{c\to0}(c/\delta)\,.
\ee
and $B'=d\widetilde B/dc|_{c=0}$.
We see that, in the logCFT, $t$ plays the role of $T$ and $b$ the role of $c$, but there are in addition logarithmic terms involving $T$.

\subsection{Two dimensions}\label{sec13}
So far we have been careful to emphasise that these results hold for CFTs in any dimension $d$.
This is a point to be stressed, since so much of the work on logCFT has focussed on two dimensions. However, since CFTs have much richer structure in $d=2$, much more can indeed be said in this case, particularly about the $c\to0$ paradox. 

For $d=2$, the spatial index structure simplifies. In complex coordinates $(z,\zb)$ the stress tensor has two independent components
$T\equiv T_{zz}$ and $\overline T\equiv T_{\bar z\bar z}$, which are holomorphic (depending only in $z$) and antiholomorphic respectively. Its 2-point functions are
\be
\langle T(z)T(0)\rangle=\frac c{2z^4}\,,\qquad\langle\overline T(z)\overline T(0)\rangle=\frac c{2\zb^4}\,,
\ee
thus defining $c$. The conformal Ward identity for primary operators implies the OPE
\be
T(z)\cdot\phi(0)=\frac{h_\phi}{z^2}\phi(0)+\frac1z\partial_z\phi(0)+\cdots\,,
\ee
and in this case (\ref{TOPE}) becomes
\begin{eqnarray}
\phi(z,\bar z)\cdot\phi(0,0)&=\frac{a_\phi}{z^{2h_\phi}\zb^{2\bar h_\phi}}
\left(1+\frac{2h_\phi}cz^2T(0)+\frac{2\bar h_\phi}c\zb^2\overline T(0)\right.\nonumber\\
&\qquad\qquad\qquad\qquad\left.+\frac{4h_\phi\bar h_\phi}{c^2}z^2
\zb^2T\overline T(0)+\cdots\right)\,,\label{OPE2}
\end{eqnarray}
where $(h_\phi,\bar h_\phi)$ are the so-called complex scaling dimensions of $\phi$, with $x_\phi=h_\phi+\bar h_\phi$ and the difference $h_\phi-\hb_\phi$ being the conformal spin of $\phi$. We have included the term proportional to $T\overline T$ in the OPE as it will play a role later. It leads to a term $\propto c^{-2}$ in the 4-point function. 

If we assume that the $c\to0$ paradox in the above is resolved in manner c) above, that is by the collision of $T$ with another operator $\widetilde T$ of scaling dimension $2+\delta$ with $\delta\to0$ as $c\to0$, \em and \em we assume this is local with respect to the other fields, then it must have conformal spin 2, the same as $T$. Therefore it has complex scaling dimensions $(2+\bar h,\bar h)$, with $\bar h=\frac12\delta$. If it is to cancel the pole at $c=0$ in the 4-point function its contribution to the OPE must have the form
\be
\big(1+O(c)\big)\frac{2h_\phi}cz^{2+\bar h}\zb^{\bar h}\widetilde T(0)\,,
\ee
where
\be
\langle\widetilde T(z,\bar z)\widetilde T(0,0)\rangle=-\frac{c(1+O(c))}{2z^{4+2\bar h}\zb^{2\bar h}}\,.
\ee
Note that we may choose the normalisation of $\widetilde T$ so that the $O(c)$ correction in the numerator is absent.
Clearly both 2-point functions vanish at $c=0$, but if we define as for general $d$
\be\label{tdef}
t\equiv\lim_{c\to0}(\widetilde T-T)/\bar h\,,
\ee
then, when $c=0$
\begin{eqnarray}
\langle t(z,\zb)t(0,0)\rangle=-\frac{2b\log(z\zb)}{z^4}\,,\label{b1}\\
\langle t(z,\zb)T(z)\rangle=\frac b{z^4}\,,\label{b2}\\
\langle T(z)T(0)\rangle=0\,,\label{b3}
\end{eqnarray}
where, with this conventional normalisation of $t$,
\be\label{bdef}
b=-\ffrac12\lim_{c\to0}(c/\bar h)\,.
\ee
Thus $(T,t)$ form a logarithmic pair. That these equations should hold in any $c=0$ CFT was postulated in 
Ref.~\cite{Gur2}
(except that the $\zb$ dependence was overlooked). One of the goals of this work is to study whether (\ref{b1}-\ref{b3}) hold in the examples we shall consider, and whether the number $b$ is a universal parameter of the logCFT. 

Similar questions may be posed in boundary logCFTs. Boundary CFT is usually considered in the upper half-plane ${\mathbb H}^+$. In this case correlation functions of boundary operators on the real axis are boundary values of analytic functions in ${\mathbb H}^+$ (except at coincident points.) In this case it makes sense to consider the analogue of (\ref{b1}--\ref{b3}) without the $\zb$ dependence, and to ask whether the boundary value of $b$ in a given logCFT is the same as in the bulk and whether it depends on the boundary conditions.

A possibly pedantic point is that (\ref{b1}) as written is dimensionally incorrect. This may be traced to the fact that in the definition
(\ref{tdef}) of $t$, the operators $T$ and $\widetilde T$ have different dimensions for $c\not=0$. Thus we should really write
\be
t\equiv\lim_{c\to0}(\widetilde T\mu^{-2\hb}-T)/\bar h\,,
\ee
where $\mu$ is the momentum scale at which the theory is renormalised. All expressions involving $\log(z\zb)$ should really be interpreted as $\log(z\zb\mu^2)$. This also applies to other logarithmic correlators, for example (\ref{e17}).
However we shall suppress this dependence in the bulk of this paper.

Notice however that even if the $c\to0$ paradox in the 4-point function is resolved in manner (a), that is by the vanishing of the normalisation $a_\phi$ faster than $c^{1/2}$, there is still a potential problem with the higher order terms in the OPE (\ref{OPE2}) proportional to $T\overline T$ and so on. For then the piece of the connected part of the 4-point function arising from 
$T\overline T$ then behaves like $a_\phi^2/c^2$. Thus, either $a_\phi$ vanishes faster than $c$, or $T\overline T$ collides with a scalar operator with dimensions $(2+\frac12\delta,2+\frac12\delta)$. This would then leave a term behaving like $a_\phi^2/c$ times a logarithm. This leaves both pieces in the connected 4-point function behaving in the same way.
If the first alternative holds, there is still a problem at higher orders, which give contributions $O(a_\phi^2/c^k)$ with $k>2$, unless they are cancelled. Logarithms are therefore unavoidable at some point.

In fact, the simplest resolution is that $a_\phi\propto c$, since this ensures that \em all \em the connected correlation functions are of the same order $O(c)$. This is what happens for the order parameter in the Potts model as $Q\to1$ (Sec.~\ref{sec4}). 

\subsection{Kac operators}

In two dimensions, the bulk operators of a non-logarithmic CFT are classified according to highest weight representations of a commuting pair of Virasoro algebras. (For boundary CFT there is just one Virasoro algebra since $\overline T$ and $T$ are not independent.) In general these are generated by all possible powers of the lowering operators $(L_n,\overline L_{\bar n})$ with $n,\bar n\leq-1$ acting on the highest weight state $|h_\phi,\bar h_\phi\rangle$. However, for rational values of the central charge $<1$ parametrised by
\be
c=1-6(p-q)^2/pq\,, 
\ee
and values of the highest weights given by the Kac formula
\begin{equation}\label{Kac}
h_{rs}=\frac{(rp-sq)^2-(p-q)^2}{4pq}\,,
\end{equation}
with $(r,s)$ positive integers, these representations contain states which are themselves of highest weight, and which should be projected out to obtain irreducible representations. We refer to operators with these highest weights as \em Kac operators\em. The fusion rules then generically dictate that in the OPE of two Kac operators labelled by $(r_1,s_1)$ and $(r_2,s_2)$ only other Kac operators with 
\be 
\fl |r_1-r_2|+1\leq r\leq r_1+r_2-1\,\quad\mbox{and}\quad |s_1-s_2|+1\leq s\leq s_1+s_2-1\,,
\ee
(and their descendants) can arise. As we shall see, this places strong constraints on possible logarithmic behaviour when a limit to a logCFT is taken. Correlation functions of Kac operators also obey linear differential equations. In the case of the 4-point function these can be reduced to ordinary equations with regular singular points where the co-ordinates coincide.

If $(p,q)$ are co-prime integers, then the Kac operators with $1\leq r\leq q$ and $1\leq s\leq p$ form a closed operator algebra. Modular invariance further constrains the physical operators of the theory to a subset of these, giving the minimal models. For $p=q+1$ these are the only unitary CFTs with $c<1$ \cite{FQS} but the other minimal models with rational $p/q$ are dense in this interval.

Another important approach to two-dimensional CFT has been through Coulomb gas methods \cite{CG}. These are particularly useful for understanding the scaling limit of lattice models which may be described in terms of planar loops, like the $O(n)$ model and $Q$=state Potts model. Although non-rigorous, these methods are very useful for extracting the scaling dimensions of the leading scaling operators of the theory. In many cases these have been shown to agree with results combining exact Bethe ansatz and finite-size scaling. The scaling dimensions found in this way fall into two groups.

In the first case they are given by the Kac formula (\ref{Kac}) for \em fixed \em values of $(r,s)$ as $n$ or $Q$ (and therefore $c$) are varied. This implies that their OPEs obey the fusion selection rules. Examples are the bulk energy operators \cite{DF} of the $O(n)$ model, with dimensions $(h_{13},h_{13})$, and of the Potts model, with dimensions $(h_{21},h_{21})$; and the boundary  $N$-leg (watermelon) operators of these models, with dimensions $h_{N+1,1}$ and $h_{1,N+1}$ respectively \cite{JCsurf,DupSal}.

The second group of operators in the Coulomb gas do not in general correspond to Kac operators except perhaps at special rational values of $c$, when they become fields of the appropriate minimal model. Examples are the bulk $N$-leg operators in the $O(n)$ and Potts models \cite{DupSal}. The decoupling of the other operators in this case arises by a collision of these operators with descendants of others. It is just this collision that may lead to logarithmic behaviour in the `extended' minimal model, that is, including the fields which formally decouple. 
Indeed, it is in this case that the indices of the differential equation obeyed by the 4-point functions may differ by integers, a situation which is well-known to lead to logarithmic solutions. 

Let us see what this tells us in general about possible resolutions of the $c\to0$ paradox. To first order in $c$ the Kac formula reads
\be\label{Oc}
h_{rs}(c)=\frac{(3r-2s)^2-1}{24}(1-c)+\frac{(3r-2s)(r-s)c}{10}+O(c^2)\,.
\ee
Thus $h_{12}(0)=0$, and for correlations of this operator the $c\to0$ paradox is resolved by choice b). In particular there are no logarithms in the 4-point function. This is the case for the boundary 1-leg operator in percolation, whose 4-point function gives the crossing formula \cite{JCcrossing}. We also see that both $h_{31}$ and $h_{15}$ are $2+O(c)$, so both are candidates for the operator $\widetilde T$ which can collide with $T$. The fusion rules dictate that the first can arise in the OPE of operators in the first row of the Kac table (corresponding to boundary $N$-leg operators of the $O(n\to0)$ model, while the second may arise in similar correlators in the $Q\to1$ Potts model. 

However, for bulk Kac operators the situation is different, for then we require $\widetilde T$ to have dimensions $(2+\delta,\delta)$, and to be a Kac operator. The only contenders for these might be $(h_{15},h_{12})$ or $(h_{31},h_{12})$,
but it is easy to see that these do not differ by 2 if $c\not=0$. The conclusion is that for bulk Kac operators with non-vanishing scaling dimensions the $c\to0$ paradox can only be resolved through choice a): that is 
\begin{center}
\em all correlations of bulk local Kac operators with $h\not=0$ must vanish at $c=0$. \em
\end{center}
In these cases, the non-zero physical correlations are usually derivatives with respect to $c$ at $c=0$. 
It should be stressed that this result does not hold for non-Kac bulk operators such as the bulk $N$-leg operators.
Neither does it hold for bulk holomorphic parafermionic Kac operators. This is because they are non-local and therefore can (and do) contain the holomorphic fields of fractional spin $(h_{31},0)$ or $(h_{15},0)$ in their OPE, which can then collide with the stress tensor.

The above no-go theorem does not hold in boundary logCFT. For then $T$ and $\overline T$ are not independent and there is no notion of conformal spin. Therefore the $c\to0$ paradox for Kac operators may be resolved by the collision of $T$ with operators with boundary scaling dimension $h_{31}$ or $h_{15}$.

Even though the correlators of bulk local Kac operators must vanish at $c=0$, if this happens by having their normalisation $a_\phi\propto c$, as discussed above, there will still be a paradox at the level of $T\overline T$ in the OPE. However
scalar operators with dimensions $(h_{31},h_{31})$ and $(h_{15},h_{15})$ \em are \em allowed by the fusion rules, and such operators then may collide with $T\overline T$ as described above. These will give to terms behaving like
$\eta^2{\bar\eta}^2\log(\eta\bar\eta)$ in the derivative of the 4-point function with respect to $c$ at $c=0$. 

\subsection{Operator products involving $t$}
Using the definition (\ref{tdef}) $t=\lim_{c\to0}\hb^{-1}(\widetilde T-T)$ and the fact that the operator $\widetilde T$ which collides with $T$ is a primary spin-2 operator, we can deduce the OPEs involving $T$ and $t$ which are usually postulated in logCFT. Begin from the standard CFT conformal Ward identities
\begin{eqnarray}
T(\ffrac12z)\cdot T(-\ffrac12z)=\frac c{2z^4}+\frac2{z^2}T(0)+\cdots\,,\\
T(\ffrac12z)\cdot\widetilde T(-\ffrac12z)=\frac{2+\bar h}{z^2}\widetilde T(0)-\frac{\hb}{2z}\partial_z\widetilde T(0)+\cdots\,,
\end{eqnarray}
together with
\be
\widetilde T(\ffrac12z)\cdot\widetilde T(-\ffrac12z)=-\frac c{2z^{4+2\hb}\zb^{2\hb}}-\frac{2+\bar h}{z^{2+2\hb}\zb^{2\hb}}T(0)+
\frac{f(\hb)}{z^{2+\hb}\zb^\hb}\widetilde T(0)+\cdots\,.
\ee
Note that we have written the symmetrised version of these equations so as to eliminate as far as possible terms involving derivatives of $T$ and $\widetilde T$. In the last equation we have chosen to normalise $\widetilde T$ so that
its 2-point function is exactly $-(c/2)/z^{4+2\hb}\zb^{2\hb}$. The second term is then fixed by associativity of the OPE. $f(\hb)$ is a universal OPE coefficient. 

Written in terms of $T$ and $t$ the last two equations become
\begin{eqnarray}
T(\ffrac12z)\cdot t(-\ffrac12z)=\frac b{z^4}+\frac2{z^2}t(0)+\frac1{z^2}T(0)-\frac1{2z}\partial_zT(0)+\cdots\,,\label{143}\\
t(\ffrac12z)\cdot t(-\ffrac12z)=-\frac{2b\log(z\zb)}{z^4}-\frac{2(\log(z\zb))^2+\log(z\zb)|+O(1)}{z^2}T(0)\nonumber\\
\quad\qquad\qquad\qquad+\frac{-4\log(z\zb)+1}{z^2}t(0)
+\cdots\,.\label{144}
\end{eqnarray}
In order for the right hand side to be finite we must have $f(\hb)=4+3\hb+\cdots$. The $O(\hb^2)$ term determines the coefficient of the $O(1)$ term multiplying $T(0)$. These equations were first written down in Ref.~\cite{GurLud}, without, however, the $\zb$-dependence.\footnote{However the non-holomorphicity of the bulk $t$ has long been recognised in the literature. See Ref.~\cite{Ridout12} for a simple derivation.} 

There is another interesting OPE expressing the non-holomorphicity of $t$.
Starting from
\be
\overline T(\zb)\cdot\widetilde T(0)=\frac{2\hb}{\zb^2}\widetilde T(0)+\frac1{\zb}\partial_\zb\widetilde T(0)+\cdots\,,
\ee
we see, writing $\widetilde T=T+\hb t$, that at $c=0$,
\be
\overline T(\zb)\cdot t(0)=\frac2{\zb^2}T(0)+\cdots\,.
\ee

We may similarly the determine the form of the OPE of $t$ with other primary fields. If $\phi$ has scaling dimensions
$(h_\phi,\hb_\phi)$ then the conformal Ward identity implies that
\be
T(z)\cdot\phi(0)=\frac{h_\phi}{z^2}\phi(0)+\frac1z\partial_z\phi(0)+\cdots\,,
\ee
while its OPE with $\widetilde T$ must have the form
\be
\widetilde T(z,\zb)\cdot\phi(0)=\frac k{z^{2+\hb}\zb^\hb}\phi(0)+\frac{k'}{z^{1+\hb}\zb^\hb}\partial_z\phi(0)+\cdots\,.
\ee
The OPE coefficients $(k,k')$ are known in principle for a given sequence of CFTs as $c\to0$, but for the OPE with $t$ to be finite we must have $k=h_\phi+O(\hb)$ and $k'=1+O(\hb)$, so that
\be\label{149}
\fl t(z,\zb)\cdot\phi(0)=-\frac{h_\phi(\log(z\zb)+O(1))}{z^2}\phi(0)-\frac{\log(z\zb)+O(1)}{z}\partial_z\phi(0)+\cdots\,.
\ee
Note that the $O(1)$ terms are universal but depend on the CFT. In Sec.~\ref{sec3} we give an example showing that they also carry a non-trivial group structure, so cannot be simply absorbed into the normalisation of the logarithmic term. 

In \em boundary \em logCFT, the main difference is that the colliding operator $\widetilde T$ has only a single scaling dimension $2+\hb$. All the above equations then hold with the dependence on $\zb$ omitted.

\subsection{Finite-size scaling and the $b$-parameter}\label{sec14}
In ordinary CFTs in $d=2$ the central charge $c$ plays a ubiquitous role. One of the most important relates to finite-size scaling of the free energy on a cylinder $S^1\times\mathbb R^1$ \cite{BCN,Aff}. Although the expectation value of the stress tensor $\langle T\rangle$ vanishes in $\mathbb R^2$  by rotational symmetry, the cylinder geometry breaks this, and in fact
\be\label{Tcyl}
\langle T\rangle=\langle\overline T\rangle= -\frac c{24}\left(\frac{2\pi}\ell\right)^2=-\frac{\pi^2 c}{6\ell^2}\,,
\ee
where $\ell$ is the circumference of $S^1$. Because $T+\overline T$ measures the response of the free energy to a change in the length $L$, this gives the finite-size correction to the free energy per unit length \cite{BCN,Aff}
\be
E=-\frac{\pi c}{6\ell}\,.
\ee
For a system with $c=0$ the free energy vanishes and a more useful quantity is the effective central charge $c'(0)$. For quenched random systems this measures the finite-size corrections to the quenched average of the free energy; for self-avoiding walks the corrections to the mean number of self-avoiding loops, and for percolation the corrections to the mean number of clusters.

However we see from (\ref{tdef}) that $\langle t\rangle$ has a finite limit at $c=0$: in fact \cite{Gur2}
\be\label{tcyl}
\langle t\rangle_{\rm cylinder}=\lim_{c\to0}\hb^{-1}\big(\langle\widetilde T\rangle-\langle T\rangle\big)
=-\frac{\pi^2b}{3\ell^2}\,.
\ee
The first term vanishes in all the examples we consider because $\widetilde T$ transforms non-trivially under ${\cal G}_n$, and we assume that this is unbroken in the vacuum state. 

A more interesting version of this result appears if we break the ${\cal G}_n$ symmetry by the boundary conditions at the end of the cylinder. In that case $\langle T\rangle$ is non-vanishing:
\be
\langle T\rangle= \left(-\frac c{24}+\Delta\right)\left(\frac{2\pi}\ell\right)^2\,,
\ee
where $(\Delta,\overline\Delta)$ are the scaling dimensions associated with the boundary state at each end. On the other hand $\langle\widetilde T\rangle$ should scale like $\ell^{-2-2\bar h}$. To get a finite limit for $\langle t\rangle$ the coefficients must cancel at $c=\bar h=0$. Hence
\be\label{tDelta}
\langle t\rangle=\Delta\lim_{\bar h\to0}\hb^{-1}\left((2\pi/\ell)^{2+2\bar h}-(2\pi/\ell)^{2}\right)\sim 
-2\Delta (2\pi/\ell)^2\log\ell\,.
\ee
We give an application of this in Sec.~\ref{sec33}.

\section{Quenched random systems and replicas}\label{sec2}
We first recall the use of the replica trick in discussing quenched random systems, and then show how their critical behaviour should be described by a logCFT. 

Consider a classical statistical mechanics system, with degrees of freedom labelled by $\{s\}$. These could be Ising spins, or something more general. In the continuum description of the pure system, let $\Psi(r)$ be some  
local operator, a functional of the $\{s\}$. The disordered system is obtained by coupling a quenched random variable $h(r)$ to $\Psi(r)$, so that the energy functional is
\be
E[\{s\}, \{h\}]=E_{\rm pure}[\{s\}]+\int h(r)\Psi(r)d^dr\,.
\ee
It is usually sufficient to assume that $h$ is white noise:
\be
\overline{h(r)}=0,\qquad\overline{h(r)h(r')}=\lambda\delta^{(d)}(r-r')\,,
\ee
where $\lambda$ measures the strength of the disorder. We use the overline to distinguish expectation values in the quenched ensemble.
For a random magnet, for example, $\Psi$ could be the local magnetisation, in which case $h$ is a random magnetic field, or $\Psi$ could be the local energy density, when $h$ corresponds to a `random $T_c$' due, for example, to non-magnetic impurities.
The quenched random variables are not dynamical, and therefore not averaged over in correlation functions, so for example, for a local observable $\Phi(r)$, its 2-point function
\be
\langle\O(r_1)\O(r_2)\rangle_{\{h\}}={\left(\frac{{\rm Tr}_s\,\O(r_1)\O(r_2)e^{-E[\{s\},\{h\}]}}
{{\rm Tr}_s\,e^{-E[\{s\},\{h\}]}}\right)}
\ee
will be different for each realisation of the quenched random variable. However, many interesting physical quantities are \em self-averaging\em, that is, for a large system as the volume $V\to\infty$ they are almost surely equal to their expectation value in the quenched random ensemble. For example
\be
\fl\lim_{V\to\infty}\int\langle\Phi(r+r')\Phi(r')\rangle_{\{h\}}\frac{d^dr'}{V}=
\overline{\langle\O(r)\O(0)\rangle}=
\overline{\left(\frac{{\rm Tr}_s\,\O(r)\O(0)e^{-E[\{s\},\{h\}]}}
{{\rm Tr}_s\,e^{-E[\{s\},\{h\}]}}\right)}\,.
\ee
This helps, for example by restoring spatial symmetries, but actually performing the quenched average is still difficult because $h$ occurs in both the numerator and the denominator $Z(\{h\})$ on the right hand side. 

There are at least two ways around this problem: the first, which works efficiently only for a gaussian ensemble, is to find some other degrees of freedom $\{\psi\}$ such that ${\rm Tr}_\psi\,e^{-E[\{\psi\},\{h\}]}=Z[\{h\}]^{-1}$, so that
\be
\overline{\langle\O(r)\O(0)\rangle}=\overline{\rm Tr}_{s,\psi}\,\O(r)\O(0)e^{-E[\{s\},\{h\}]-E[\{\psi\},\{h\}]}\,.
\ee
Performing the quenched average is now easy. In some cases this leads to a \em supersymmetry \em between $\{s\}$ and $\{\psi\}$.
We mention this here because this case applies to disordered non-interacting electrons or waves, when the Green's function can be written as a gaussian functional integral. After quenched averaging, the models become effectively interacting, and, at a critical point should correspond to a CFT with $c=0$, since the total partition function is unity. 
However, we will not pursue this method here, partly because, as we shall show, more physical information can be gained by considering the behaviour as $c\to0$ rather than exactly at $c=0$. 

The second method introduces $n$ copies of the original degrees of freedom $\{s_a\}$, where $a\in(1,\ldots,n)$,
called replicas, and uses the fact that, for positive integer $n$,
\be
\overline{\langle\O(r)\O(0)\rangle}=\overline{\left(\frac{{\rm Tr}_{s_a}\,\O_1(r)\O_1(0)e^{-\sum_{a=1}^nE[\{s_a\},\{h\}]}}{Z[\{h\}]^n}\right)}\,.
\ee
If this can then be continued to $n=0$, so that the denominator is unity, then the quenched average is straightforward. This however then couples the replicas. At the critical point, we assume that this is described by a CFT, with an internal symmetry group $S_n$. For weak randomness this replica trick is just a way of organising the weak disorder expansion in $\lambda$. However in other examples, where non-perturbative effects are important, its use may be suspect, and this should always be borne in mind, especially if non-physical results seem to emerge.

\subsection{Replica group theory and OPEs}
The symmetry group is $S_n$ and, for $\lambda=0$, it acts trivially on the non-interacting replicas. For $\lambda\not=0$, we assume that, at the critical point, the theory flows under the RG to a CFT which, for generic $n$, is non-logarithmic (although irrational except perhaps for special values of $n$), with scaling operators transforming according to irreducible representations of $S_n$. In particular, the multiplet $(\O_1,\ldots,\O_n)$ decomposes into 
\begin{eqnarray}
\O\equiv\sum_{a=1}^n\O_n\,,\\
\widetilde\O_a\equiv\O_a-(1/n)\sum_{a=1}^n\O_a\,.
\end{eqnarray}

The basic tool we shall use is the OPE. We assume that the overall structure of the OPE in the interacting theory and the pure theory is the same, consistent with the $S_n$ symmetry, and that only the OPE coefficients and scaling dimensions are modified. That is, we assume that the interacting theory is a \em deformation \em of the non-interacting one. This, of course, is a standard principle of quantum field theory, and does not imply the convergence of perturbation theory. It can be justified in an RG setting, if, for example, there is some upper critical dimension at which the two fixed points collide. 

In the non-interacting theory we have the OPE (suppressing indices)
\be
\O_a(r)\cdot\O_b(0)=\delta_{ab}r^{-2x_{\rm pure}}\left(1+\cdots+B\frac{x_{\rm pure}}{c_{\rm pure}}r^dT_a(r_1)+\cdots\right)\,.
\ee
Here $T_a$ is the stress tensor of the $a$th replica. 
In terms of the irreducible operators $\widetilde\O_a$ and $\O$ these become
\begin{eqnarray}
\widetilde\O_a\cdot\widetilde\O_b=(\delta_{ab}-\frac1n)r^{-2x_{\rm pure}}\left(1+B\frac{x_{\rm pure}}{nc_{\rm pure}}
r^dT\right)\nonumber\\
\qquad\qquad\qquad\qquad+\left(B\frac{x_{\rm pure}}{c_{\rm pure}} r^d(\delta_{ab}\widetilde T_a-\frac1n\widetilde T_a-\frac1n
\widetilde T_b)+\cdots\right)\,,\\
\widetilde\O_a\cdot\O=\frac1nr^{-2x_{\rm pure}}B\frac{x_{\rm pure}}{nc_{\rm pure}}\widetilde T_a+\cdots\,,\\
\O\cdot\O=n|r_{12}|^{-2x_{\rm pure}}\left(1+B\frac{x_{\rm pure}}{nc_{\rm pure}}
r_{12}^dT+\cdots\right)\,,
\end{eqnarray}
where $T$ and $\widetilde T_a$ are the corresponding irreducible linear combinations of the $T_a$. Note that the stress tensor of the combined system appears with the correct coefficient reflecting its central charge $nc_{\rm pure}$.

In the interacting theory these deform to 
\begin{eqnarray}
\widetilde\O_a\cdot\widetilde\O_b=\widetilde A(n)(\delta_{ab}-\frac1n)r^{-2\tilde x(n)}\left(1+B\frac{\tilde x(n)}{c(n)}
r^dT\right)\nonumber\\
\qquad\qquad+\left(\widetilde B(n)\frac{n\tilde x(n)}{c(n)} r^{d+\delta(n)}(\delta_{ab}\widetilde T_a-\frac1n\widetilde T_a-\frac1n
\widetilde T_b)+\cdots\right)\,,\label{21}\\
\widetilde\O_a\cdot\O=\frac1nr^{x(n)+\tilde x(n)}\hat B(n)\frac{x(n)}{c(n)}r^{d+\delta(n)}\widetilde T_a+\cdots\,,\label{22}\\
\O\cdot\O=nA(n)|r_{12}|^{-2x(n)}\left(1+B\frac{x(n)}{c(n)}
r_{12}^dT+\cdots\right)\,,\label{23}
\end{eqnarray}
where $(x(n),\tilde x(n),d,d+\delta(n))$ are the dimensions of $(\O,\widetilde\O_a,T,\widetilde T_a)$ respectively,
 and $c(n)$ is the central charge of the interacting theory.
Note that $T=\sum_aT_a$ is the stress tensor of the replicated theory. When the replicas interact, its dimension $d$ is protected, but that of $\widetilde T_a$ will in general change. Also, the coefficient $B$ of $T$ in the above OPEs is again fixed by conformal symmetry in the interacting theory, but  $(\widetilde B(n),\hat B(n)) $ are not. However they should approach $B$ as $n\to0$. Similarly $A(n)$ and $\widetilde A(n)$
should approach unity.

If we look at the 2-point functions of the original replicated observables we have
\begin{eqnarray}
\langle\O_1(r)\O_1(0)\rangle=(1-\frac1n)\frac{\widetilde A(n)}{r^{2\tilde x(n)}}+\frac1n\frac{A(n)}{r^{2x(n)}}\,,\\
\langle\O_1(r)\O(0)\rangle=\frac{A(n)}{r^{2x(n)}}\,,\\
\langle\O(r)\O(0)\rangle=\frac {nA(n)}{r^{2x(n)}}\,.
\end{eqnarray}

But
\begin{eqnarray}
\lim_{n\to0}\langle\O_1(r)\O_1(0)\rangle=\overline{\langle\O(r)\rangle\langle\O(0)\rangle}\,,\\
\lim_{n\to0}\langle\O_1(r)\O(0)\rangle=\overline{\langle\O(r)\O(0)\rangle-\langle\O(r)\rangle\langle\O(0)\rangle}\,,
\end{eqnarray}
and these had better be finite, since the right hand sides are the quenched averages of the disconnected and connected 2-point functions respectively. Therefore, as $n\to0$, we must have $\widetilde A(n)-A(n)\to0$ and $\tilde x_\O(n)-x_\O(n)\to0$, so that
\begin{eqnarray}
\lim_{n\to0}\langle\O_1(r)\O_1(0)\rangle\sim\frac{2A(0)\big(\tilde x'(0)-x'(0)\big)\log r}{r^{2x(0)}}\,,\\
\lim_{n\to0}\langle\O_1(r)\O(0)\rangle\sim\frac{A(0)}{r^{2x(0)}}\,.
\end{eqnarray}
Note that the logarithmic behaviour occurs only in the quenched average of the disconnected part of the correlation function. Since the connected part is given by functional derivatives of the free energy, this is consistent with the latter having non-logarithmic scaling near the critical point. Also note that ratio of the coefficient of the logarithmic correlator to the non-logarithmic one is universal. 

$(\O_1,\O)$ are an example of a logarithmic pair. However, in this example the rank of the Jordan cell containing these fields is really $\big(1+(n-1)\big)$, as $n\to0$, at least in terms of bosonic fields. 

Of course, we only find the logarithm above if $\tilde x_\O(n)-x_\O(n)\to0$ and $\tilde x'_\O(0)-x'_\O(0)$ is finite. The first condition can be justified as follows. In $d=2$ the partition function on the torus $S^1\times S^1$ is the generating function for all the scaling dimensions of a conventional (non-logarithmic) CFT: $Z\propto\sum_jd_jq^{x_j}$, where $q$ the modular parameter and $d_j$ is the degeneracy of the scaling dimension $x_j$. In higher dimensions this extends to the partition function on $S^{d-1}\times S^1$ \cite{JChigherd}. Applying this to the interacting replicated theory for $n\to0$, when $Z\to1$ we see that the scaling dimensions with $x_j\not=0$ must arrange themselves so that their contribution cancels in this limit. In this case, the $(n-1)$ independent operators $\widetilde\O_a$ must cancel the contribution of $\O$, so that the total contribution to $Z$ is
\be
Z=q^{x_\O(n)}+(n-1)q^{\tilde x_{\O}(n)}+\mbox{other powers of $q$}\,.
\ee
The two terms can cancel only if $\tilde x_\O(n)-x_\O(n)\to0$. That the derivative with respect to $n$ is non-zero may be seen in perturbation theory in the disorder strength. The coupling between the replicas has the form $\lambda\int\Delta(r)d^dr$ where
\be
\Delta=\sum_{a\not=b}\Psi_s(r)\Psi_b(r)\,.
\ee
The shift in the scaling dimension of an operator $\phi$ in $k$-th order perturbation theory may be computed from knowledge of the correlation functions 
\be
\langle\phi(r_1)\phi(r_2)\rangle_k\equiv\langle\phi(r_1)\phi(r_2)\left(\int\Delta(r)d^dr\right)^k\rangle_{\rm pure}
\ee
in the unperturbed CFT, as long as the operator $\phi$ is normalised correctly. Taking $\phi=\widetilde\O_a$ and $\O$ in turn, and using the $S_n$ symmetry, we find after some algebra
\begin{eqnarray}
\langle\widetilde\O_a(r_1)\widetilde\O_a(r_2)\rangle_k
\propto\langle\O_1(r_1)\O_1(r_2)\rangle_k-\langle\O_1(r_1)\O_2(r_2)\rangle_k\,,\label{pp1}\\
\langle\O(r_1)\O(r_2)\rangle_k
\propto\langle\O_1(r_1)\O_1(r_2)\rangle_k+(n-1)\langle\O_1(r_1)\O_2(r_2)\rangle_k\,.\label{pp2}
\end{eqnarray}
Note that these agree when $n=0$ but in general are different because the interaction couples the replicas so 
$\langle\O_1(r_1)\O_2(r_2))\rangle_k\not=0$. This can be worked out explicitly \cite{Cardy1} when $\Phi$ is taken to be $\Psi$, the operator to which the disorder couples, for then there is a contribution at first order $k=1$. 

However there are at least two provisos to the above discussion. The first is when the symmetry group is larger than $S_n$. For example, in the random-bond Ising model in zero external field the symmetry group is ${\mathbb Z}_2^n\otimes S_n$, that is within each replica we can take $s_a\to-s_a$. In that case if $\Phi=s$, the second term on the right hand side of (\ref{pp1},\ref{pp2}) vanishes by this symmetry, so there are no logarithms in the 2-point spin-spin correlations, only in the energy-energy correlations. Another way to state this is to observe that $s_a$ transforms irreducibly under the larger group. 

The second proviso is that the whole discussion assumes that the limit $n\to0$ is smooth in the RG equations. This is the case for disorder coupling to the local energy in a ferromagnet, so would apply to cases where this is relevant by the Harris criterion \cite{Harris}, for example the three dimensional random-bond Ising model or the random-bond $Q$-state Potts model in two dimensions for $Q>2$ \cite{Ludwig}. However for the random field Ising model, it is known that the RG is singular at $n=0$ \cite{Brezin}. 

Finally, from the OPEs (\ref{21}-\ref{23}) we can see how the $c\to0$ paradox is evaded in this theory. The $\Phi\cdot\Phi$ OPE does this by having the normalisation of $\Phi$ vanish as $n\to0$. In (\ref{21}) we should recall that because
$\langle TT\rangle=O(c)$ while $\langle \widetilde T_a\widetilde T_b\rangle=(\delta_{ab}-(1/n))\times O(c)$, the terms involving $T$ and $\widetilde T$ are both of the same order. If we work out the OPE $\O_a\cdot\O_b$ of the \em physical \em fields we find the terms involving $T$ and $\widetilde T$ to be, as $n\to0$
\be
\delta_{ab}r^{-2x(0)}\left(B\frac{x(0)}{c(n)}r^dT+\widetilde B(n)\frac{nx(0)}{c(n)}r^{d+\delta(n)}\widetilde T_a\right)\,.
\ee
Since (supressing index structure)
\be
\langle T(r)T(0)\rangle=\frac{c(n)}{r^{2d}}\,,
\qquad \langle\widetilde T_a(r)\widetilde T_b(0)\rangle=\left(\delta_{ab}-\frac1n\right)\frac{c(n)}{r^{2(d+\delta)}}\,,
\ee
we see that indeed $\widetilde T_a$ plays the role of $\widetilde T$ in resolution (c) of the paradox.
However note that if we define $t$ as before actually transforms reducibly under $S_n$, unlike $T$, which is a singlet.

\subsection{Higher rank operators}
In the above we assumed that the operators in the replicated theory carry a single replica index. But there will, in addition, be higher rank objects which carry more indices. These should arise, for example, in the OPE between different replicas
which has the form
\be
\widetilde\O_a\cdot\widetilde\O_b=\widetilde A(n)\left((\delta_{ab}-1/n){\mathbf 1}+\O^{(2)}_{ab}+\cdots\right)\,,
\ee
where $\sum_a\O^{(2)}_{ab}=0$. 
The correlators of $\O^{(2)}_{ab}$ measure the correlations between correlation functions in different realisations of the disordered system \cite{Davis}, for example
\be
\langle\O^{(2)}_{12}(r)\O^{(2)}_{12}(0)\rangle=\overline{\langle\O(r)\O(0)\rangle^2}\,.
\ee
In spin glasses, when $\O$ is the local magnetisation, the quenched average correlations $\overline{\langle\O(r)\O(0)\rangle}$ vanish, and it is the above quantities which indicate the onset of spin glass order. In this example
$\O^{(2)}_{ab}$ is known as the Edwards-Anderson order parameter \cite{EA}, usually denoted by $q_{ab}$.

Note that in the above $\O^{(2)}_{ab}$ satisfies $\sum_a\O^{(2)}_{ab}=0$, so that only the components with $a\not=b$ are independent. However these do not transform irreducibly:
$\sum_{a,b(a\not=b)}\O^{(2)}_{ab}$ transforms a singlet, and the sums of each row or column $\sum_{b(b\not=a)}\O^{(2)}_{ab}$
transform the same way that $\O_a$ itself does. Hence we may define irreducible combinations\footnote{These coincide, up to unimportant differences, with those found by Vasseur \etal\cite{V1} in their study of the Potts model.}
\begin{eqnarray}
\O^{(2)}\equiv\sum_{a,b(a\not=b)}\O^{(2)}_{ab}\,,\label{n1}\\
\widetilde\O^{(2)}_a\equiv\sum_{b(\not=a)}\O^{(2)}_{ab}-\frac1n\O^{(2)}\,,\label{n2}\\
\hat\O^{(2)}_{ab}\equiv\delta_{a\not=b}\left(\O^{(2)}_{ab}-\frac1{n-2}(\widetilde\O^{(2)}_a+\widetilde\O^{(2)}_b)
-\frac1{n(n-1)}\O^{(2)}\right)\,.\label{n3}
\end{eqnarray}
The coefficients in the last line are determined by demanding $\sum_{b(b\not=a)}\hat\O^{(2)}_{ab}=0$. It may be checked that this implies that 2-point correlators between $\hat\O^{(2)}_{ab}$ and the other representations vanish, as expected if it is irreducible. 

The form of the the non-zero 2-point functions is also fixed by the $S_n$ group to be \cite{V1}
\begin{eqnarray}
\langle\O^{(2)}(r)\O^{(2)}(0)\rangle&=2n(n-1)A^{(2)}(n)\,r^{-2x^{(2)}(n)}\,,\label{c1}\\
\langle\widetilde\O^{(2)}_a(r)\widetilde\O^{(2)}_b(0)\rangle&=(n-2)\widetilde A^{(2)}(n)\left(\delta_{ab}-\frac1n\right)\,r^{-2\tilde x^{(2)}(n)}\,,\label{c2}\\
\langle\hat\O^{(2)}_{ab}(r)\hat\O^{(2)}_{cd}(0)\rangle&=\hat A^{(2)}(n)
\left(\delta_{ac}\delta_{bd}+\delta_{ad}\delta_{bc}-\frac1{n-2}\big(\delta_{ac}+\delta_{ad}+\delta_{bc}+\delta_{bd}\big)\right.\nonumber\\
&\qquad\qquad\qquad\left.+\frac2{(n-1)(n-2)}\right)\,r^{-2\hat x^{(2)}(n)}\,,\label{c3}
\end{eqnarray}
where the prefactors are chosen so that the amplitudes are unity in the non-interacting case.

From this we see that the physical correlator as $n\to0$ (picking out only the terms which are potentially singular in this limit) is
\begin{eqnarray}
\fl\langle\Phi^{(2)}_{ab}(r)\Phi^{(2)}_{cd}(0)\rangle&\sim \frac14\langle(\widetilde\O^{(2)}_a(r)+\widetilde\O^{(2)}_b(r))
(\widetilde\O^{(2)}_c(0)+\widetilde\O^{(2)}_d(0))\rangle+\frac1{n^2}\langle\O^{(2)}(r)\O^{(2)}(0)\rangle\\
&\sim\frac2n\widetilde A^{(2)}(n)r^{-2\tilde x^{(2)}(n)}-\frac2nA^{(2)}(n)r^{-2x^{(2)}(n)}\\
&\sim 4A^{(2)}(0)\big(x^{(2)}{}'(0)-\tilde x^{(2)}{}'(0)\big)\,r^{-2x^{(2)}(0)}\log r\,.
\end{eqnarray}
Note that this logarithmic behaviour occurs in correlations between all pairs of replicas. For the spin glass this corresponds to the disconnected part $\overline{\langle q(r)\rangle\langle q(0)\rangle}$.

\section{The $O(n)$ model}\label{sec3}
In general dimension $d$, the conventional $O(n)$ field theory is based on a multiplet of $n$ commuting fields $\phi_a(r)$ with $a\in(1,\ldots,n)$ and action
\be
S=\int\left[\ffrac12\sum_a(\partial\phi_a)^2+\ffrac12m_0^2\sum_a\phi_a^2+\lambda\sum_{a,b}\phi_a^2\phi_b^2\right]d^dr\,.
\ee
If this theory is regularised on the lattice, and the quartic term is undone by a gaussian transformation, the partition function corresponds to an ensemble of closed loops, each weighted by a fugacity $n$, with an on-site interaction, repulsive if $\lambda>0$. In this realisation $n$ can be any real number, although for non-negative weights $n\geq0$. 
Correlators such as $\langle\phi_a(r_1)\phi_a(r_2)\rangle$ correspond to open paths connecting $r_1$ and $r_2$ in the background of these closed loops. In particular taking $n=0$ gives an ensemble of self-repelling walks \cite{deGennes} which model linear polymers in solution.  For $\lambda\to+\infty$ these become self-avoiding. At the critical point the mean loop length diverges and the model becomes critical, described in the scaling limit by a CFT which is in general non-unitary. For $n=0$ the partition function $Z=1$ and we expect this to become a logCFT.

Let us examine how this happens at the level of the OPE. First consider the case of an operator, such as $\phi_a$ itself, which transforms according to the $n$-dimensional (vector) representation of $O(n)$. Note that, unlike the case of $S_n$, this is irreducible. We may think of the interacting theory as a deformation of the gaussian theory with $\lambda=0$. In the non-interacting theory the OPE takes the form (suppressing indices)
\be
\phi_a(r)\cdot\phi_b(0)=a_\phi r^{-2x_\phi}\left(\delta_{ab}+r^{2x_\phi}:\!\phi_a\phi_b\!:+B\frac{x_\phi}{c_{\rm free}}r^dT_{ab}+\cdots\right)\,,
\ee
where $T_{ab}^{\mu\nu}=\partial^\mu\phi_a\partial^{\nu}\phi_b+\cdots$. The diagonal components $T_{aa}$ are the stress tensor of each copy in the non-interacting theory.

Before turning on the interactions, the operators :$\phi_a\phi_b$: and $T_{ab}$ should be decomposed into their irreducible parts
\begin{eqnarray}
\phi^{(2)}\equiv\sum_a:\!\phi_a^2\!:\,,\qquad
\tilde\phi^{(2)}_{ab}\equiv:\!\phi_a\phi_b\!:-(1/n)\delta_{ab}\phi^{(2)}\label{phi2}\,;\\
T^{\mu\nu}\equiv\sum_aT^{\mu\nu}_{aa}\,,\qquad
\widetilde T^{\mu\nu}_{ab}\equiv T^{\mu\nu}_{ab}-(1/n)\delta_{ab}T^{\mu\nu}\,.
\end{eqnarray}
$\phi^{(2)}$ is the energy operator of the theory, coupling to the bare mass $m_0^2$, and $\tilde\phi^{(2)}_{ab}$ for
$a\not=b$ is the 2-leg operator, acting as a source for mutually avoiding pairs of walks.

On switching on the interactions the OPE deforms into 
\begin{eqnarray}\label{37}
\fl\phi_a\cdot\phi_b=a_\phi r^{-2x_\phi(n)}\left(\delta_{ab}+(f/n)\delta_{ab}r^{x^{(2)}(n)}\phi^{(2)}
+\tilde fr^{\tilde x^{(2)}(n)}\tilde\phi^{(2)}_{ab}\right.\nonumber\\
\qquad\qquad\qquad\left.+B\delta_{ab}\frac{x_\phi(n)}{c(n)}r^dT+
\widetilde B(n)\frac{x_\phi(n)}{c(n)}r^{d+\delta(n)}\widetilde T_{ab}+\cdots\right)\,,\label{OnOPE}
\end{eqnarray}
where $x^{(2)}(n)$ and $\tilde x^{(2)}(n)$ are the scaling dimensions of $\phi^{(2)}$ and $\tilde\phi^{(2)}_{ab}$, and the OPE coefficients $f,\tilde f$ and $\widetilde B$ are expected to have finite limits as $n\to0$, just as they do in the free theory. 
Since the 2-point function $\langle\phi_a(r_1)\phi_a(r_2)\rangle$ counts walks from $r_1$ to $r_2$, the coefficient $a_\phi(n)$ should have a finite limit as $n\to0$. Once again, $B$ is fixed by conformal symmetry. 

Notice that because $\phi_a$ transforms irreducibly, there are now no logarithms in its  2-point function, unlike the previous section.
However, the 2-leg operator is logarithmic as $n\to0$. The form of the 2-point functions above is fixed by the $O(n)$ symmetry to be
\begin{eqnarray}
\langle\tilde\phi^{(2)}_{ab}(r)\tilde\phi^{(2)}_{cd}(0)\rangle=\widetilde A(n)(\delta_{ac}\delta_{bd}+\delta_{ad}\delta_{bc}-(2/n)
\delta_{ab}\delta_{cd})r^{-2\tilde x^{(2)}(n)}\,,\label{38}\\
\langle\phi^{(2)}(r)\phi^{(2)}(0)\rangle=2nA(n)r^{-2x^{(2)}(n)}\,,\label{Onenergy}
\end{eqnarray}
where the coefficients $A(n),\widetilde A(n)$ should have finite limits as $n\to0$ as in the non-interacting theory. 
It can be checked \cite{Cardy1} that, to first order in $\lambda$ (equivalent to the $\epsilon$-expansion), the scaling dimensions 
$x^{(2)}(n)$ and $\tilde x^{(2)}(n)$ are different except at $n=0$. The above correlators are singular as $n\to0$, but the 2-point function of the physical field $\phi^{(2)}_{ab}=\tilde\phi^{(2)}_{ab}
+(1/n)\phi^{(2)}\delta_{ab}$ is finite:
\be
\fl\langle\phi^{(2)}_{ab}(r)\phi^{(2)}_{cd}(0)\rangle\sim r^{-2x^{(2)}(0)}
\left[\delta_{ab}\delta_{cd}({x^{(2)}}{}'(0)-{\tilde x^{(2)}}{}'(0))
\log r+\delta_{ac}\delta_{bd}+\delta_{ad}\delta_{bc}\right]\,.
\ee
The physical interpretation of this is as follows. In the language of paths, $\phi^{(2)}_{ab}(r)$ is the 2-leg operator which creates two mutually avoiding self-avoiding walks at $r$, carrying `colours' $a$ and $b$. Similarly $\phi^{(2)}_{cd}(0)$ creates a pairs of walks  at $0$ with colours $c$ and $d$. They can  join up only if the colours agree. Thus the first, logarithmic term, corresponds to a pair of mutually avoiding closed loops, one passing through $r$ and the other through  $0$.  On the other hand, the non-logarithmic term describes a pair of walks from $0$ to $r$. As before, we see that the logarithmic behaviour arises in the disconnected piece of the correlation function.

It is straightforward to see from (\ref{OnOPE}) how the $c\to0$ paradox is avoided in this case: it is resolved by the collision of
$T$ with $\widetilde T_{ab}$. If we normalise these operators so that (suppressing spatial index structure)
\begin{eqnarray}
\langle T(r)T(0)\rangle=c(n)/r^{2d}\,,\\
\langle\widetilde T_{ab}(r)\widetilde T_{cd}(0)\rangle=-c(n)\big(\delta_{ab}\delta_{cd}-(n/2)(\delta_{ac}\delta_{bd}+\delta_{ad}\delta_{bc})\big)/r^{2d+2\delta(n)}\,,\label{312}
\end{eqnarray}
(the index structure being fixed as in (\ref{38})) then as long as $\widetilde B(n)\to B$, the paradox in (\ref{37}) is resolved.
It would be an interesting exercise to compute $\delta(n)$ in the $\epsilon$-expansion. However it is easy to see that it vanishes at $n=0$ as it should.

 \subsection{Higher rank operators}
 
 It is interesting to study possible logarithmic behaviour in correlators of higher composite operators. For example, consider the 4-leg operator $\phi^{(4)}_{abcd}$ which occurs in the symmetrised OPE of four fields $\phi_a$. In this case the irreducible operators are 
 \begin{eqnarray}
 \phi^{(4)}&=\sum_{a,b}\phi^{(4)}_{aabb}\,,\\
 \tilde\phi^{(4)}_{ab}&=\sum_c\phi^{(4)}_{abcc}-(1/n)\phi^{(4)}\delta_{ab}\,,\\
 \hat\phi^{(4)}_{abcd}&=\phi^{(4)}_{abcd}-\frac1{n+4}\left(\tilde\phi^{(4)}_{ab}\delta_{cd}+\mbox{5 other permutations}\right)\nonumber\\
 &\qquad\quad\,\,-\frac1{n(n+2)}\phi^{(4)}\left(\delta_{ab}\delta_{cd}+\mbox{2 other permutations}\right)\,.
 \end{eqnarray}
 Thus the theory is also potentially logarithmic at $n=-2$ and $-4$.  It can be seen from the index structure above that logarithms occur only in pieces of the 2-point functions corresponding to walks which begin and end at the same point. The case $n=-2$ is interesting because the loop corrections to the 2-point function of $\phi_a$ vanish \cite{minus2}. In $d=2$ it corresponds to a logCFT with $c=-2$.

\subsection{Two dimensions}
In two dimensions, the scaling dimensions of the various operators above are known exactly.
In the bulk theory they are most easily found using Coulomb gas methods \cite{CG}, which we describe briefly. In two dimensions, the weight $n$ for each loop may be written in terms of local, if complex, weights by orienting each loop and assigning a weight $e^{\pm i\chi\theta/2\pi}$ to each time a loop turns right-(left-)wards through an angle $\theta$. The correct weight per loop is recovered if $n=e^{i\chi}+e^{-i\chi}= 2\cos\chi$. The ensemble of oriented loops may be mapped to a model of heights $h\in\pi{\mathbb Z}$ on the dual lattice, where the unit current along each loop is the lattice curl of $h$. The measure on $h$ is then local, and it is assumed that in the scaling limit this becomes a free gaussian field with action
\be
S=(g/4\pi)\int(\nabla h)^2d^2r\,.
\ee
There are various ways of determining $g$, with the result $g=1-\chi/\pi$. Here $1\leq g\leq 2$ corresponding to the critical point of the dilute phase of the $O(n)$ model, and $\frac12\leq g\leq 1$ the dense phase. The central charge $c$ and scaling dimensions are most easily found by studying the theory on a cylinder of circumference $\ell$ and length $L\gg\ell$. In general such a free field theory has $c=1$ so that the partition function goes like $Z\sim\exp(\pi L/6\ell)$. However the above prescription miscounts loops which wrap around the cylinder. Correcting this changes the partition function to $\exp(\pi cL/6\ell)$ where
\be
c=1-6(g-1)^2/g\,.
\ee
$n=0$ in the dilute phase corresponds to $g=\frac32$ which gives $c=0$ as expected. The scaling dimensions of the $N$-leg operator $\tilde\phi^{(N)}$
are found by requiring that (at least) $N$ open paths traverse the cylinder from end to end, by imposing a discontinuity of $N\pi$ in $h$ around the cylinder. Thus we write $h=\pi Nv/\ell+\tilde h$, where $v$ is a coordinate around the cylinder, and $\tilde h$ is periodic. The action for the deterministic piece is $(g/4\pi)(N\pi/\ell)^2\ell L$, so that the correlation function behaves like
\be
\fl\langle\phi^{(N)}(L)\phi^{(N)}(0)\rangle\sim\exp\left(-(g/4\pi)(N\pi)^2(L/\ell)+(\pi L/6\ell)-(\pi cL/6\ell)\right)\,.
\ee
The second term in the exponent comes from the integral over the fluctuations of $\tilde h$, and the last from dividing by $Z$. This gives the $N$-leg bulk exponent \cite{DupSal}
\be\label{Nleg}
\tilde x^{(N)}(n)=\frac{gN^2}8-\frac{(g-1)^2}{2g}\quad\mbox{where}\quad n=-2\cos(\pi g)\,.
\ee
One the other hand, the singlet energy operator $\phi^{(2)}$ has scaling dimension $x^{(2)}(n)=2h_{13}$, that is, it is a Kac operator \cite{DF}. This is consistent with its 2-point function vanishing at $n=0$ as can be seen from (\ref{Onenergy}). At $n=0$ we see that $x^{(2)}(0)=\tilde x^{(2)}(0)=\frac23$. However, their derivatives with respect to $n$ at $n=0$ are different, so we see that $\tilde\phi^{(2)}$ is indeed the logarithmic partner of $\phi^{(2)}$ at $n=0$. 

In order to identify $\widetilde T_{ab}$, the logarithmic partner of the stress tensor, which has non-zero spin, we need first to twist the field $h$ by some amount $v_0$ along the cylinder so that $h(L,v)=h(0,v-v_0)$. Thus we now write $h=-(\pi Nv_0/\ell)(u/L)+\pi Nv/\ell+\tilde h$ where $u$ is a coordinate running along the cylinder. This gives an extra factor $\exp(-(g/4\pi)(\pi Nv_0/\ell L)^2(\ell L))$. In order to give something of spin $\sigma$ we should multiply by $e^{i(2\pi/\ell)\sigma v_0}$ and integrate over $v_0$. This gives $\exp(-(4\pi/gN^2)\sigma^2(L/\ell))$.

Thus the spin $\sigma$ $N$-leg operator has scaling dimension
\be\label{spin}
\tilde x^{(N)}_\sigma(n)=\tilde x^{(N)}(n)+\frac2{gN^2}\sigma^2\,.
\ee
Taking $N=2$ and $\sigma=2$ we then find\footnote{This Coulomb gas calculation was first reported in \cite{V2}.}
\be
 x_{\widetilde T}=\tilde x^{(2)}+\frac2g=1+\frac3{2g}\,.
\ee
 For $n=0$, $g=\frac32$, we see that $x_{\widetilde T}=2$ as expected. In this case (\ref{bdef}) then gives
\be
 b_{\rm bulk}(\mbox{SAW})=-\frac1{dx_{\widetilde T}/dc|_{c=0}}=-5\,.
\ee
 
 On the other hand, on the boundary there is no concept of conformal spin. The energy operator is the stress tensor and the operator which collides with this at $c=0$ is the 2-leg operator, which has scaling dimension $h_{31}$. From (\ref{Oc}) we see that $dh_{31}/dc|_{c=0}=-\frac35$ so that
\be
 b_{\rm boundary}(\mbox{SAW})=\ffrac56\,.
\ee
However we note that this value could depend on the boundary conditions, since the boundary operator content for $c\not=0$, and therefore the possible identification of $\widetilde T$, may well do so.

\subsection{The $t$ operator}\label{sec33}
It is interesting to see how the general arguments of Sec.~\ref{sec1} work in this example. If we define $t$ as in (\ref{tdef}) then we must recognise that in fact it carries $O(n)$ indices:
\be
t_{ab}=
\lim_{c\to0}\hb^{-1}\big(\widetilde T_{ab}-T\delta_{ab}\big)\,.
\ee
In this case the 2-point function (\ref{b1}) becomes, using the 2-dimensional version of (\ref{312})
\be\label{323}
\langle t_{ab}(z,\zb)t_{cd}(0)\rangle=-\frac{2b\log(z\zb)}{z^4}\delta_{ab}\delta_{cd}+\frac\beta{z^4}(\delta_{ac}\delta_{bd}+\delta_{ad}\delta_{bc})\,,
\ee
where $\beta=\lim_{c\to0}(nc/2\hb^2)$. Note that this second term is universal, and cannot simply be absorbed into a change of scale of the logarithm since it has a different index structure. 

It is possible also to determine the correct version of the further terms in the OPEs (\ref{143}, \ref{144}). We find that the logarithmic terms involve only the singlet part $\sum_at_{aa}$ of $t_{ab}$ and so are consistent with the assumption that it transforms this way. However the non-logarithmic terms carry a non-trivial $O(n)$ index structure, with universal coefficients. We stress that these terms are physically observable in higher-order correlation functions, because the group structure gives information about connectedness in the language of curves.

Similarly the $O(1)$ terms in the OPE of $t_{ab}$ with a general operator $\phi$, see (\ref{149}), will depend on the representation content of $\phi$, and, unless it transforms as a singlet, will carry non-trivial index structure.
On the cylinder with vacuum states at either end, $\langle\widetilde T_{ab}\rangle=0$ so $\langle t_{ab}\rangle\propto\delta_{ab}$ with same coefficient as in (\ref{tcyl}). 
However if we introduce states $\langle\phi_c|$ and $|\phi_d\rangle$ at either end
\be
\langle\phi_c|t_{ab}|\phi_d\rangle=\lim_{c\to0}\hb^{-1}\left(\langle\phi_c|\widetilde T_{ab}|\phi_d\rangle
-\delta_{ab}\langle\phi_c|T|\phi_d\rangle\right)\,.
\ee
 The second term is  given by (\ref{Tcyl}) as before, multiplied by $\delta_{cd}$. The first term is proportional to 
 $\delta_{ab}\delta_{cd}-(n/2)(\delta_{ac}\delta_{bd}
 +\delta_{ad}\delta_{bc})$ and scales like $\ell^{-2-2\bar h}$. Therefore (\ref{tDelta}) becomes
 \be\label{325}
 \langle\phi_c|t_{ab}|\phi_d\rangle\sim-2\Delta(2\pi/\ell)^2\left(\delta_{ab}\delta_{cd}\log\ell +\gamma(\delta_{ac}\delta_{bd}
 +\delta_{ad}\delta_{bc})\right)\,,
 \ee
 where $\gamma=\lim_{c\to0}(n/4\hb)$.
 Thus, as far as its logarithmic behaviour goes, $t_{ab}$ behaves like an $O(n)$ singlet, and it is meaningful to discuss a single operator $t$. However, the non-logarithmic corrections again reveal its more complicated group theoretic properties.
 
 The physical interpretation of these two terms is as follows.  For an SAW, the expectation value of stress tensor $\langle  T(z)\rangle$ is proportional to the spin-2 moment of the probability that the curve $\Gamma$ intersects the line segment $L(z,\epsilon,\theta)$ connecting $z\pm\epsilon e^{i\theta}$ \cite{Doy}:
 \be
 T(z)=\lim_{\epsilon\to0}(8/\pi\epsilon^2)\int_0^{2\pi}P(\Gamma\cap L(z,\epsilon,\theta)\not=\emptyset)\,e^{-2i\theta}d\theta\,.
 \ee
 On a cylinder this is non-vanishing because of the spatial anisotropy provided by the axis of the cylinder. When the vacuum state is at both ends then $\Gamma$ is a closed loop passing through $z$. When the fields are placed at either end, there is an open curve traversing the whole cylinder. The index structure in (\ref{325}) shows that in the non-logarithmic term $\Gamma$ is this curve, in the absence of any closed loops. In the logarithmic term, however, $\Gamma$ is a closed loop in the presence of the open curve.

\section{The $Q$-state Potts model and percolation}\label{sec4}
The $Q$-state Potts model is a generalisation of the Ising model. On a lattice the state at each site is labelled by a random variable $s(r)$ which takes $Q$ possible values, $a\in(1,\cdots,Q)$. The unnormalised Boltzmann weights are given by a product over bonds $\prod_{rr'}e^{K_{rr'}\delta_{s(r)s(r')}}$, with $K_{rr'}\geq0$ in the ferromagnetic case. By writing this as proportional to $\prod_{rr'}\big((1-p_{rr'})+p_{rr'}\delta_{s(r)s(r')}\big)$, with $p_{rr'}=1-e^{-K_{rr'}}$, and expanding out the product, the partition function may be interpreted as a sum over configurations of random clusters. In this model the bond $(rr')$ is open with probability $p_{rr'}$ and closed with probability $1-p_{rr'}$.  Clusters are subsets of sites connected by open bonds. On summing over the $s(r)$, each cluster carries a weight $Q$. In this way the model is also defined when $Q$ is not a positive integer. There is a critical point at $p=p_c$ at which, for example, the mean cluster size diverges. For sufficiently large $Q>Q_c$ this transition is first order, but for smaller values of $Q$ it has a scaling limit described by a CFT which is in general non-unitary. The case $Q=1$ corresponds to bond percolation, for which the partition function $Z=1$, so in this case we expect the scaling limit to be a logCFT, although other values of $Q$ ($-2$, corresponding to uniform spanning trees, and $+2$, corresponding to the extended Ising model) also turn out to be logarithmic. 

The symmetry group of the Potts model is $S_Q$. As we shall argue, the singular nature of its representation  theory at $Q=1$ can be used to identify logarithmic operators. The identification of its logarithmic operator content has been carried out in detail by Vasseur \etal\cite{V1}. Here we adopt a slightly different field-theoretic approach, with, nevertheless, the same conclusions.

The Potts model may also be realised as an $Q$-component ferromagnet  with cubic symmetry in an external field along the $(1,1,\ldots,1)$ axis. If $S_a$ is the local magnetisation, the cubic symmetry restricts it to taking the values
$(\pm1,\pm1,\ldots,\pm1)$. The external field gives rise to a non-zero expectation $\langle S_a\rangle$, which is independent of $a$, along the $(1,1,\ldots,1)$ axis, but the energetically favoured fluctuations are into the $Q$ possible states
where locally $\sum_{a=1}^QS_a=Q-1$. These can be interpreted as the states of an $S_Q$-symmetric Potts model. 
The order parameter
\begin{equation}\label{constraint}
\phi_a\equiv S_a-(1/Q)\sum_aS_a\quad\mbox{satisfies}\quad\sum_{a=1}^Q\phi_a=0\,.
\end{equation}
The lowest order terms in the effective field theory describing these fluctuations allowed by the $S_Q$ symmetry are then
\be
\int[\sum_a(\partial\phi_a)^2+m_0^2\sum_a\phi_a^2+\lambda\sum_a\phi_a^3+O(\phi^4)]d^dr\,,
\ee
with, however, the constraint (\ref{constraint}).
Although the cubic terms imply a first order transition in mean field theory for integer $Q\geq3$, in the RG analysis near the upper critical dimension $d=6$ there is an accessible fixed point describing a continuous transition for $Q<2$ \cite{PottsRG}. In two dimensions, the fluctuations drive the transition to be continuous for $Q\leq4$.

In the analysis of possible logarithmic terms arising within the OPE, the constraint (\ref{constraint}) is crucial, since it implies that the fields $\{\phi_a\}$ transform according to the \em irreducible \em $(Q-1)$-dimensional representation of $S_Q$. This is what makes this example different from that quenched random case in Sec.~\ref{sec2}, even though the symmetry group is the same. It means that the 2-point function $\langle\phi_a(r)\phi_b(0)\rangle$ is \em not \em expected to exhibit logarithmic behaviour. 

Instead, it is necessary to consider composite operators which transform according to higher representations of $S_Q$. In particular consider the operators arising in the the OPE, schematically written
\be\label{pp}
\phi_a\cdot\phi_b=a_\phi\big(\delta_{ab}-(1/Q)\big){\mathbf 1}+\phi^{(2)}_{ab}+\cdots\,.
\ee
The first term is proportional to the identity operator and its form is fixed by the requirement that $\sum_a\phi_a=0$.
In percolation, the probability that two sites $(r_1,r_2)$ are in the same cluster is given by 
\be
P(r_1,r_2)=\lim_{Q\to1}(Q-1)^{-1}\langle\phi_a(r_1)\phi_a(r_2)\rangle\,.
\ee

In the second term of (\ref{pp}) the leading behaviour is subtracted off to obtain $\phi^{(2)}_{ab}\equiv$:$\phi_a\phi_b$:. This is symmetric under $a\leftrightarrow b$ and also satisfies
$\sum_a\phi^{(2)}_{ab}=\sum_b\phi^{(2)}_{ab}=0$. This may be decomposed into operators which transform irreducibly under $S_Q$ exactly as in Sec.~\ref{sec2}, so the analysis there maybe taken over \em mutatis mutandis \em with the substitutions $\Phi\to\phi$ and $n\to Q$. Therefore we do not repeat the relevant formulae here. However for percolation the relevant limit is $Q\to1$. From (\ref{n1}-\ref{n3}) we see that the 2-point function of $\hat\phi^{(2)}_{ab}$ becomes singular there. However the physical operator is (keeping only the leading terms as $Q\to1$)
\be
\phi^{(2)}_{ab}\sim \hat \phi^{(2)}_{ab}+\frac1{Q-1}\phi^{(2)}\,,
\ee
with a 2-point function
\begin{eqnarray}
\langle \phi^{(2)}_{ab}(r)\phi^{(2)}_{cd}(0)\rangle&\sim
-\frac{2\hat A^{(2)}(Q)}{Q-1}\,r^{-2\hat x^{(2)}(Q)}+\frac{2A^{(2)}(Q)}{Q-1}\,r^{-2x^{(2)}(Q)}\nonumber\\
&\sim4A^{(2)}(1)(\tilde x^{(2)}{}'(1)-x^{(2)}(1))\,r^{-2x^{2)}(1)}\log r\,.
\end{eqnarray}

The correlations of the energy operator $\phi^{(2)}$ in percolation have been identified with those of the red bonds, that is those bonds whose opening or closing either connect or disconnect two neighbouring large clusters. However the logarithmic correlations are seen to occur once again in the disconnected parts, that is, terms in which for example $(ab)\not=(cd)$.
In this case the red bonds are constrained to connect/disconnect \em different \em pairs of clusters \cite{V1}.

In the non-interacting theory, the stress tensor $T^{\mu\nu}=\sum_a\partial^\mu\phi_a+\partial^\nu\phi_a+\cdots$ has companions $T^{\mu\nu}_{ab}=\sum_a\partial^\mu\phi_a+\partial^\nu\phi_b+\cdots$ which should deform and decompose under the interactions into irreducible components $T$, $\widetilde T_a$ and $\hat T_{ab}$ just as for the 2-cluster operator above.

\subsection{The logarithmic Ising model and uniform spanning trees}
We also see from (\ref{n3}) that the correlations of $\phi^{(2)}_{ab}$ also become logarithmic at $Q=2$. This is the random cluster representation of the Ising model. It contains more observables, of a geometric nature, than the conventional unitary Ising CFT. From (\ref{c1}-\ref{c3}) we see that (keeping now the most singular terms as $Q\to2$)
\begin{eqnarray}
\langle \phi^{(2)}_{ab}(r)\phi^{(2)}_{cd}(0)\rangle&\sim
\frac{\hat A^{(2)}(Q)}{Q-2}(2-\delta_{ac}-\delta_{ad}-\delta_{bc}-\delta_{bd})\,r^{-2\hat x^{(2)}(Q)}\nonumber\\
&\quad+\frac{\tilde A^{(2)}(Q)}{Q-2}(\delta_{ac}+\delta_{ad}+\delta_{bc}+\delta_{bd}-(4/Q))\,r^{-2\tilde x^{(2)}(Q)}\\
&\sim(\delta_{ac}+\delta_{ad}+\delta_{bc}+\delta_{bd}-2)\, r^{-2x^{(2)}(2)}\,\log r\,,
\end{eqnarray}
where $x^{(2)}(2)$ is the scaling dimension of the usual energy operator in the Ising model.

Since $a\not=b$ and $c\not=d$, it would appear that when $Q=2$ the only possibilities are that each of these pairs is either $(12)$ or $(21)$, in which case the prefactor above always vanishes. However this is not correct: for example we may choose $a,b,c$ and $d$ to be all different before continuing to $Q=2$. This would again relate to the probability that there are four non-overlapping FK clusters, two of them touching at $r$ and two at $0$. Once again, the logarithm arises in this disconnected piece. 

The limit $Q\to0$ in the Potts model forces all the sites to be in the same cluster. When in addition $p\to0$ this minimises the number of open bonds, so the cluster has the structure of a spanning tree. In this case the order parameter $\phi_a$ is not logarithmic, because it transforms irreducibly. But we see from (\ref{n1},\ref{n2}) that the 2-cluster operator $\tilde\phi^{(2)}_{a}$ is. From (\ref{c1},\ref{c2}) we see that in this limit 
\be
\left\langle\sum_{b(\not=a)}\phi^{(2)}_{ab}(r)\sum_{d(\not=c)}\phi^{(2)}_{cd}(0)\right\rangle\sim r^{-2x^{(2)}(0)}\log r\,.
\ee
The operator $\sum_{b(\not=a)}\phi^{(2)}_{ab}(r)$ forces there to be another cluster in the system, touching the first cluster at $r$. We note that the log CFT with $c=-2$ corresponding to uniform spanning trees is also related to that for dense polymers, which has been well-studied using a fermionic description \cite{cminus2} and it would be interesting to compare the two approaches.

By considering higher rank operators in the Potts model, it should be possible to exhibit operators which become logarithmic at other values of $Q$. Since the group theory factors should be rational functions of $Q$ we expect these to occur at algebraic values of $Q$ in general. However in two dimensions it may well be that there are in general higher degeneracies enforced by the quantum group structure which modify the simple considerations based on $S_Q$.

\subsection{Two dimensions}
Once again the various scaling dimensions introduced above are known exactly in two dimensions. 
In the Coulomb gas approach, the closed loops now refer to the hulls which separate clusters and dual clusters. At the critical point each loop is counted with a weight $\sqrt Q$. Thus we now have $\sqrt Q=-2\cos\pi g$, and the Potts model turns out to correspond to the dense phase of the $O(n)$ model, with $\frac12\leq g\leq1$. In particular, percolation corresponds to
$g=\frac23$. 

Note that the 2-cluster operator $\hat\phi^{(2)}_{ab}$ is related to a 4-leg operator since the clusters are separated by dual clusters. From (\ref{Nleg}) we see that $\tilde x^{(2)}=\frac54$. On the other hand the energy operator $\phi^{(2)}$ is a Kac operator \cite{DF} with scaling dimension $2h_{21}=\frac54$.
Thus they collide at $Q=1$ as expected. However it is easy to check that their derivatives with respect to $Q$ do not agree, and therefore $\hat\phi^{(2)}$ is the logarithmic partner of the energy operator. This is also true on the boundary, where the energy operator is the stress tensor with dimension 2, and $\hat\phi^{(2)}$ is the boundary 4-leg operator with scaling dimension $h_{15}=2$ at $Q=1$. 

The scaling dimension of the bulk companion to the stress tensor may be read off from (\ref{spin}) setting $N=4$ and $\sigma=2$. This gives \cite{V2}
\be
x_{\widetilde T}=\tilde x^{(2)}+\frac1{2g}=1+\frac{3g}2\,.
\ee
The important point here is that, since $c(g)=c(1/g)$, then $\delta=x_{\widetilde T}-2$ is the \em same \em function of $c$ as in the $O(n)$ model. Hence \cite{V2}
\be
b_{\rm bulk}(\mbox{percolation})=b_{\rm bulk}(\mbox{SAW})=-5\,.
\ee
On the other hand, from (\ref{Oc}) we see that $dh_{15}/dc|_{c=0}=\frac45$, so
\be
b_{\rm boundary}(\mbox{percolation})=-\ffrac58\,.
\ee
However we again remark that this may depend on the boundary conditions. For example it is known \cite{JCsurf} that in the Potts model the boundary condition changing operator $\phi_{(a|\not=a)}$ is a Kac operator of dimension $h_{21}$.
In order to resolve the $c\to0$ paradox in the 4-point function of this operator $\widetilde T$ must have dimension $h_{31}$, and so $b_{\rm boundary}=\frac56$ in this case.

\section{On the uniqueness of $b$}\label{sec5}
The resolution (c) of the $c\to0$ paradox discussed in Sec.~\ref{sec1} postulates the existence of an operator $\widetilde T$ of scaling dimension $d+\delta$ which collides with $T$, see (\ref{OPE2}). If this is the case, we may extract the 
$b$-parameter as
$b=-(dc/d\delta)|_{c=0}$. We have seen examples of this in Secs.~(\ref{sec2}--\ref{sec4}). However, from this point of view, it may happen that \em different \em operators $\widetilde T$ perform this function, depending on which OPE is being considered. This might lead to different values of $b$.

There is a simple argument why this should not be the case. For suppose we consider two different operators $\phi_1$ and $\phi_2$.  Their respective OPEs take the form
\begin{eqnarray}
\phi_1(r)\cdot\phi_1(0)=\frac{a_{\phi_1}}{r^{2x_{\phi_1}}}\left(1+B\frac{x_{\phi_1}}cr^dT+\widetilde B\frac{x_{\phi_1}}cr^{d+\delta_1}\widetilde T_1+\cdots\right)\,,\\
\phi_2(r)\cdot\phi_2(0)=\frac{a_{\phi_2}}{r^{2x_{\phi_2}}}\left(1+B\frac{x_{\phi_2}}cr^dT+\widetilde B\frac{x_{\phi_2}}cr^{d+\delta_2}\widetilde T_2+\cdots\right)\,.
\end{eqnarray}
They both couple to the same stress tensor $T$ but to different colliding operators $\widetilde T_1$ and $\widetilde T_2$.
Both $a_{\phi_1}$ and $a_{\phi_2}$ are assumed to be non-vanishing.

However, if $\delta_1\not=\delta_2$ then conformal invariance implies that the 2-point function $\langle\widetilde T_1\widetilde T_2\rangle$ vanishes. If we then consider the 4-point function $\langle\phi_1(r_1)\phi_1(r_2)\phi_2(r_3)\phi_2(r_4)\rangle$ in the limit when $r_{12}\sim r_{34}\ll r_{13}\sim r_{24}$, we see that the pole at $c=0$ is not cancelled. To avoid this $\delta_1(c)=\delta_2(c)$, and so $b$ is the same in both cases.

This may used to explain why the same value $b\!=\!-5$ appears in both bulk percolation and bulk self-avoiding walks.
Although they have the same value of $c$, they correspond to different values of $g$ ($\frac23$ and $\frac32$) in the Coulomb gas. In order to use the above argument we need to show that operators corresponding to $g=\frac32$ occur in bulk percolation. We may do this by appealing to duality in the description of these models using Schramm-Loewner evolution (SLE) \cite{SLE}. In this approach, conformally invariant models of planar random curves are parametrised by $\kappa=4/g$. In models with $\kappa<4$ ($g>1$) these curves are simple, while for $\kappa>4$ ($g<1$) they have double points. However, when $\kappa>4$ one may define the exterior boundary of such a curve, which turns out to be an SLE with parameter $16/\kappa$. This SLE duality \cite{SLEduality} is between models with the same value of $c$. We see then that we can define operators $\phi_1$ in percolation and $\phi_2$ in SAW, such that the
4-point function $\langle\phi_1\phi_1\phi_2\phi_2\rangle$ is non-zero. Therefore they must have the same value of $b_{\rm bulk}$. Note that does not imply that \em every \em logCFT with $c=0$ has $b_{\rm bulk}=-5$.

In systems with a boundary, the above argument fails if $r_1,r_2$ are on the boundary and $r_3,r_4$ in the bulk, for then
$\langle \widetilde T_1(r_1)\widetilde T_2(r_3)\rangle$ does not have to vanish even when they have different scaling dimensions. This is because the 2-point function with one argument on the boundary behaves under conformal mappings like a bulk 3-point function, with the third argument being the reflection of the bulk point in the boundary. Thus it is quite possible to have \em different \em values of $b$ in the bulk and at the boundary, in the same logCFT. This is what happens in percolation and self-avoiding walks.

Purely within the boundary theory, however, the same argument as above should hold and therefore the boundary value of $b$ should be unique. Why, then, are the boundary values of $b$ different for percolation and self-avoiding walks?
In the first case, $b$ is related to the derivative of $h_{15}$ with respect to $c$, in the second it is the derivative of $h_{31}$. The fusion rules dictate that $h_{15}$ occurs only in the OPEs of operators with scaling dimensions $h_{1,N+1}$.
These are the dimensions of the boundary $N$-leg operators in the Potts model. Similarly, $h_{31}$ occurs in the OPEs of the $N$-leg boundary operators in the $O(n)$ model, with scaling dimensions $h_{N+1,1}$. Thus the paradox is resolved if the former set of operators, corresponding to the first column of the Kac table, does not arise in the boundary $O(n)$ model, and the second set, corresponding to the first row, does not arise in the boundary $Q$-state Potts model.\footnote{Except that $h_{12}=0$ may occur in a theory containing operators in the first row, since this operator evades the $c\to0$ paradox by choice (b).}


\section{Torus partition function}\label{sec6}
In two dimensions, the partition function on a torus encodes the scaling dimensions $(h,\bar h)$ of an ordinary CFT \cite{JCtorus}:
\begin{equation}\label{Ztorus}
Z=(q\bar q)^{-c/24}\sum_{h,\bar h}d(h,\bar h)q^h{\bar q}^{\bar h}\,,
\end{equation}
where $d(h,\bar h)$ is the degeneracy of operators with dimensions $(h,\bar h)$ (a positive integer for a unitary CFT.)
Here $q=e^{2\pi i\tau}$ is the modular parameter of the torus constructed by identifying opposite edges of a parallelogram with vertices at $(0,1,\tau,1+\tau)$. $Z$ has the property of modular invariance under $\tau\to-1/\tau$ and $\tau\to\tau+1$, which is highly non-trivial and constrains the operator content. 

Let us see what taking the limit as $n\to n_c$ tells us about the torus partition function for a logCFT. For a theory with $c=0$, $Z=1$, and therefore it is the derivative with respect to $n$ which is physical. (For quenched random systems it is the quenched free energy, for SAW it is the partition function for a single closed loop, for percolation it is the mean cluster number.) Let us first consider the $O(n)$ model. The first few terms in the partition function for $n\not=0$ are
\begin{eqnarray}
Z(n)=(q\bar q)^{-c(n)/24}\left(1+n|q|^{x^{(1)}(n)}+|q|^{x^{(2)}(n)}+\ffrac12(n-1)(n+2)|q|^{\tilde x^{(2)}(n)}\right.\nonumber\\
\qquad\qquad\left. +q^2+\ffrac12(n-1)(n+2)q^{2+\delta(n)}{\bar q}^{\delta(n)}+\cdots \right)\,.
\end{eqnarray}
The first term corresponds to the identity operator, the next is the 1-leg operator, and so on. We have decomposed the 2-leg operator as in (\ref{phi2}). The $q^2$ term is the contribution of the stress tensor, and then that of the symmetric traceless tensor $\widetilde T_{ab}$ which collides with it at $n=0$. If we differentiate this with respect to $n$, the leading contribution is $-(c'(0)/12)\log|q|$. The full expression has two sets of terms: a non-logarithmic piece from differentiating the degeneracy factors, and a piece $\propto\log|q|$ times a sum of powers of $q$ and $\bar q$. In this example this takes the form
\be
Z_{\rm log}'(0)=-(c'(0)/12)\log|q|\left(1+\frac{12}{\tilde b^{(2)}}|q|^{x^{(2)}(0)}+
\frac{12}{b}q^2+\cdots\right)\,.
\ee
where $\tilde b^{(2)}=\lim_{c\to0}c(n)/(x^{(2)}(n)-\tilde x^{(2)}(n))$.
This now encodes the \em logarithmic \em operator content of the theory, and the coefficient of $q^h{\bar q}^{\bar h}$ is the universal coefficient of the ratio of the amplitudes of the $\langle DD\rangle$ correlator and $\langle CD\rangle$. In particular the coefficient of  $q^2$ is proportional to the inverse of the $b$-parameter. 

A similar thing happens in the Potts model as $Q\to1$:
\begin{eqnarray}
\fl Z(Q)=(q\bar q)^{-c(Q)/24}\left(1+(Q-1)|q|^{x^{(1)}(Q)}+|q|^{x^{(2)}(Q)}+(Q-1)|q|^{\tilde x^{(2)}(Q)}\right.\nonumber\\
\fl\qquad\left.+\ffrac12Q(Q-3)|q|^{\hat x^{(2)}(Q)}
q^2+(Q-1)q^{2+\tilde\delta(Q)}{\bar q}^{\tilde\delta(Q)}+\ffrac12Q(Q-3)q^{2+\hat\delta(Q)}{\bar q}^{\hat\delta(Q)}+\cdots\right)
\end{eqnarray}
Note that the term $\propto q^{2+\tilde\delta(Q)}{\bar q}^{\tilde\delta(Q)}$ does not contribute to $Z_{\rm log}$ at $Q=1$, so that the coefficient of $q^2$ depends only on $b$ as in the previous example.

At $Q=2$, the partition function $Z(Q=2)$ collapses into that of the unitary Ising model. In order to expose logarithmic operators it is necessary to consider the derivative $\partial Z(Q)/\partial Q|_{Q=2}$.

\section{Non-critical logarithmic field theories}\label{sec7}
A CFT deformed by a relevant operator flows under the RG to another fixed point. The theory develops a mass or length scale, beyond which it crosses over to another RG fixed point described by another CFT, or, more typically, to a trivial fixed point describing the infrared behaviour of a massive quantum field theory. It is this case we consider here, when the ultraviolet CFT is a logCFT which is a limit of a conventional CFT as described above.

For $n\not=n_c$ there are two possible scenarios: either the relevant operator preserves symmetry under ${\cal G}_n$, or it breaks it. In the first case, the massive states of the QFT will transform according to irreducible representations of ${\cal G}_n$, and the the second case any subgroup which remains. We shall consider only the first case here.

As an example take the general case of $S_n$ symmetry considered above. The spectrum of the massive theory will in general depend on the details of the dynamics, but one scenario is that there is a single excitation with mass $m(n)$, transforming according to the singlet representation of $S_n$, and an $(n-1)$-plet with masses $\tilde m(n)$. Just as for the scaling dimensions $x(n)$ and $\tilde x(n)$, we expect these to be different for $n\not=0$ but to become degenerate at $n=0$. If we now consider the correlation functions $\langle\Phi(r)\Phi(0)\rangle$ and $\langle\widetilde\Phi_a(r)\widetilde\Phi_b(0)\rangle$ and insert a complete set of states, then in the first case only singlet states can contribute and in the second only states transforming according to the $(n-1)$-dimensional representation. Thus (\ref{32}) is replaced by
\be\label{F}
\fl\langle\widetilde\Phi_a(r)\widetilde\Phi_b(0)\rangle=\frac{(\delta_{ab}-1/n)\,\widetilde F_n\big(\tilde m(n)r\big)}{r^{2\tilde x(n)}}\,,\qquad
\langle\Phi(r)\Phi(0)\rangle=\frac{n\,F_n\big(m(n)r\big)}{r^{2x(n)}}\,,
\ee
Here $F_n$ and $\widetilde F_n$ are scaling functions, expected to have finite limits as $n\to0$. Note that the cross-correlator $\langle\widetilde\Phi_a\Phi\rangle$ still vanishes because they transform according to different representations. 

The 2-point functions of $\Phi_a$  and $\Phi_b$ with $a\not=b$ are however
\begin{eqnarray}
\langle\Phi_a(r)\Phi_a(0)\rangle=\frac{(1-1/n)\widetilde F_n\big(\tilde m(n)r\big)}{r^{2\tilde x(n)}}+\frac1{n}\frac{F_n\big(m(n)r\big)}{r^{2x(n)}}\,,\label{pf1}\\
\langle\Phi_a(r)\Phi_b(0)\rangle=-\frac1n\frac{\widetilde F_n\big(\tilde m(n)r\big)}{r^{2\tilde x(n)}}+\frac1{n}\frac{F_n\big(m(n)r\big)}{r^{2x(n)}}\,.\label{pf2}
\end{eqnarray}
For these to be finite we must therefore have $\tilde m(n)-m(n)\to0$ and $\widetilde F_n-F_n\to0$. The leading behaviour is seen to be 
\be
\langle\Phi_a(r)\Phi_a(0)\rangle\sim\langle\Phi_a(r)\Phi_b(0)\rangle\sim r\big(m'(0)-\tilde m'(0)\big)\frac{F_0'\big(m(0)r\big)}{r^{2x(0)}}\,.
\ee
In particular, if for general $n$ the correlators decay $\propto e^{-m(n)r}$, at $n=0$ the logarithmic pieces gain an extra factor of $r$, corresponding to a double pole in momentum space. 

\subsection{Logarithmic sum rules}
In two dimensions, Zamoldochikov's $c$-theorem \cite{Zam} states that there is a function $C$ defined on the space of all unitary renormalised QFTs which is decreasing along RG trajectories and stationary at fixed points, when it takes the value of the central charge $c$ of the corresponding CFT. 

Let us briefly recall its original derivation \cite{Zam}. Away from a fixed point, the trace $\Theta=T^\mu_\mu$ of the stress tensor is non-zero. However rotational invariance is preserved, so we identify the spin $(2,0,-2)$ components $(T,\Theta,\overline T)$.
Rotational symmetry implies that their 2-point functions in ${\mathbb R}^2$ have the form
\begin{eqnarray}
\langle T(z,\zb)T(0,0)\rangle=F(z\zb)/z^4\,,\\
\langle \Theta(z,\zb)T(0,0)\rangle=G(z\zb)/z^3\zb\,,\\
\langle\Theta(z,\zb)\Theta(0,0)\rangle=H(z\zb)/z^2\zb^2\,.
\end{eqnarray}
The conservation equation $\partial_{\zb}T+\ffrac14\partial_z\Theta=0$ then implies
\begin{eqnarray}
\dot F+\ffrac14(\dot G-3G)=0\,,\label{Z1}\\
\dot G-G+\ffrac14(\dot H-2H)=0\,,\label{Z2}
\end{eqnarray}
where $\dot F=z\zb F'(z\zb)$, \em etc. \em Defining $C=2F-G-\frac38H$, we see that
\be
\dot C=-\ffrac34H\,.
\ee
In a unitary QFT, $H\geq0$, which shows that $C$ is a decreasing function of $r^2=z\zb$. For $r\to0$ and $\infty$ it equals the central charge of the respective UV or IR CFT. Although in the non-unitary CFTs we consider in this article it is no longer necessarily true that $H\geq0$, an integrated version, known as the $c$-theorem sum rule \cite{JCsumrule}, still holds:
\be
c_{\rm UV}-c_{\rm IR}=\frac34\int_0^\infty \big(H(r^2)/r^2\big)d(r^2)=\frac3{4\pi}\int r^2\langle\Theta(r)\Theta(0)\rangle_c\,d^2r\,.
\ee

For a relevant perturbation $\lambda\int\phi\,d^2r$ by an operator with scaling dimension $x_\phi<2$,
\be
\Theta(r)=-2\pi\lambda(2-x_\phi)\phi(r)
\ee
\cite{Zam}, and so
\be\label{sr1}
c_{\rm UV}-c_{\rm IR}=3\pi\lambda^2(2-x_\phi)^2\int r^2\langle\phi(r)\phi(0)\rangle_c\,d^2r\,.
\ee

We also note that this is not the only sum rule derivable from (\ref{Z1},\ref{Z2}). From the first equation alone we see that
$2\dot F+\frac12\dot G=\frac32G$, so that
\be\label{sr2}
c_{\rm UV}-c_{\rm IR}=-\frac32\int_0^\infty \big(G(r^2)/r^2\big)d(r^2)=-\frac3{2\pi}\int z^2\langle\Theta(r)T(0)\rangle_c\,d^2r\,.
\ee

Let us apply these to the case of a logarithmic pair $(C,D)$ in a logCFT with $c=0$, which is a limit as $n\to 0$ of an ordinary CFT. 
Recalling (\ref{CDdef}) we have
\be
C=\big(x(n)-\tilde x(n)\big)\phi\,,\quad D=\phi-\tilde\phi\,.
\ee
We assume that the CFT is perturbed by the operator $\phi\propto C$. This is because, in the examples we have considered, the non-logarithmic companion $C$ always transforms as a singlet under ${\cal G}_n$, and we want to have the symmetry preserved in the massive theory. Thus we consider a perturbation $\lambda\int C(r)d^2r$ of the action.
Since the connected correlation functions of $C$ are all $O(c)$, this is the correct normalisation so that the perturbed free energy remains $O(c)$. Using 
$\Theta(r)=-2\pi\lambda(2-x(n))C(r)$, we then have from (\ref{sr1})
\begin{eqnarray}
c(n)=3\pi\lambda^2(2-x(n))^2\int r^2\langle C(r)C(0)\rangle d^2r\\
= 3\pi\lambda^2(2-x(n))^2(x(n)-\tilde x(n))\int r^2\langle C(r)D(0)\rangle d^2r\,.
\end{eqnarray}
The $\langle CD\rangle$ correlator is finite as $n\to0$ and thus we have the sum rule in the logarithmic theory
\be
3\pi\lambda^2(2-x)^2\int r^2\langle C(r)D(0)\rangle d^2r=\tilde b=\frac{c'(0)}{x'(0)-\tilde x'(0)}\,.
\ee

We can go further and consider the integral of the logarithmic correlator
\begin{eqnarray}
\int r^2\langle D(r)D(0)\rangle d^2r=\int r^2\langle\phi(r)\phi(0)\rangle d^2r+\int r^2\langle\tilde\phi(r)\tilde\phi(0)\rangle d^2r\\
=(x(n)-\tilde x(n))^{-1}\int r^2\langle C(r)D(0)\rangle d^2r+\int r^2\langle\tilde\phi(r)\tilde\phi(0)\rangle d^2r\,.
\end{eqnarray}
We have assumed that the cross-correlator $\langle\phi\tilde\phi\rangle$ vanishes. This true in all our examples because $\phi$ and $\tilde\phi$ transform according to different irreducible representations of ${\cal G}_n$. 

The first term  on the right hand side is known from (\ref{sr1}) and is $O(1/c)$. Hence, for the left hand side to be finite, the second term on the right hand side must have a singular piece of the form
\be
-\frac{c'(0)}{3\pi\lambda^2(2-x)^2(x'(0)-\tilde x'(0))(x(n)-\tilde x(n))}\,\xi^{2(x(n)-\tilde x(n))}\,,
\ee
where $\xi=m^{-1}$ is the correlation length. This comes from the scaling form (\ref{F}) of the $\langle\tilde\phi\tilde\phi\rangle$ correlator. As discussed previouslyThe  ultraviolet cut-off $\Lambda$ appears to get the overall dimensions correct, and arises from the fact that $\phi$ and $\tilde\phi$ renormalise differently. 

Putting together the two terms and taking the limit, we then find the logarithmic $c$-theorem sum rule
\be\label{722}
3\pi\lambda^2(2-x)^2\int r^2\langle D(r)D(0)\rangle d^2r=-2\tilde b\log\xi+O(1)\,.
\ee
The indeterminacy in the $O(1)$ term reflects the indeterminacy in adding a multiple of $C$ to $D$.
As discussed previously, the  renormalisation scale $\mu$ should appear multiplying $\xi$ to get the overall dimension correct, arising from the fact that $\phi$ and $\tilde\phi$ renormalise differently.

Let us now consider the second sum rule (\ref{sr2}). For a perturbation by $C$ this reads, for $n\not=0$,
\be\label{unint}
c(n)=3\lambda(2-x(n))\int z^2\langle C(r)T(0)\rangle d^2r\,.
\ee
This is not very interesting, but now consider its logarithmic analogue, replacing $T$ by $t=(\widetilde T-T)/\hb$, see (\ref{tdef}). Then, for $c\not=0$,
\be
\fl\int z^2\langle C(r)t(0)\rangle d^2r=\hb^{-1}\int z^2\langle C(r)\widetilde T(0)\rangle d^2r
-\hb^{-1}\int z^2\langle C(r)T(0)\rangle d^2r\,.
\ee
The first term vanishes in all the examples we have considered since $C$ and $\widetilde T$ transform according to different irreducible representations of ${\cal G}_n$. Thus, using (\ref{bdef}) we have the sum rule at $c=0$
\be
2b=3\lambda(2-x)\int z^2\langle C(r)t(0)\rangle d^2r\,.
\ee
We can go further, replacing $C$ by $D$. Then
\be
\fl\int z^2\langle D(r)t(0)\rangle d^2r=-\hb^{-1}\int z^2\langle \phi(r)T(0)\rangle d^2r-\hb^{-1}\int z^2\langle\tilde\phi(r)\widetilde T(0)\rangle d^2r\,,
\ee
where again the cross-terms vanish by ${\cal G}_n$ symmetry. The first term on the right hand side is proportional to the right hand side of (\ref{unint}), and is $O(1/c)$.  Therefore the second term, which scales relative to the first like 
$\xi^{x(n)-\tilde x(n)-2\bar h(n)}$, must cancel this. This gives the sum rule for the correlator of the logarithmic operators
\be\label{728}
3\lambda(2-x)\int z^2\langle D(r)t(0)\rangle d^2r=-2(b+\tilde b)\log\xi+O(1)\,.
\ee
As discussed in Sec.~\ref{sec3}, the $O(1)$ terms in (\ref{722},\ref{728}) should actually carry a group index structure, as should $D$ and $t$. 

We remark that since $\widetilde T$ and $\widetilde \Theta=-2\pi(2-x)D$ are not components of a conserved current for $c\not=0$
we cannot use an equation analogous to (\ref{Z2}). Thus is it does not seem possible in this class of theories to derive the type of $b$-theorem discussed by Gurarie \cite{Gur2} for supersymmetric logCFTs.

\subsection{The logarithmic $\Delta$-sum rule}
There is an even simpler sum rule holding in any perturbed CFT: let $\Phi$ be an operator which gains a non-zero expectation in the perturbed theory, which for the purposes here, we assume to be massive. Then
\be
x_\Phi\langle\Phi\rangle=-S_d^{-1}\int \langle\Theta(r)\Phi(0)\rangle_c\,d^dr\,,
\ee
where $S_d$ is the volume of the unit sphere $S^{d-1}$, conventionally included in the definition of the stress tensor ($S_2=2\pi$.) This follows from the identification of the trace $\Theta$ of the stress tensor as the generator of scale transformations. A more careful derivation for $d=2$ is given in Ref.~\cite{Del}. It assumes that the integral converges at short distances, otherwise a subtraction is required. 

Applied to the case where $C$ is the perturbing operator, we have $\Theta=-S_d(d-x(n))C$, so
\be
x\langle C\rangle=\lambda(d-x(n))\int \langle C(r)C(0)\rangle_c d^dr\,,
\ee
as long as $x(n)<d/2$. Both sides are $O(c)$ as $c\to0$, because they are both derivatives of the free energy. 
We may write this as
\be
\lambda(d-x)\int \langle C(r)D(0)\rangle_c\,d^dr=x\langle D\rangle\,.
\ee
By a similar argument to the above we may also write
\begin{eqnarray}
\fl\int \langle D(r)D(0)\rangle_c\,d^dr
=\frac{1}{(x(n)-\tilde x(n))^{2}}\int \langle C(r)C(0)\rangle_c\,d^dr+\int \langle\tilde\phi(r)\tilde\phi(0)\rangle_c\,d^dr\\
=\frac{x\langle C\rangle}{\lambda(d-x(n))(x(n)-\tilde x(n))^{2}}+\lambda(d-x(n))\int \langle\tilde\phi(r)\tilde\phi(0)\rangle_c\,d^dr\,.
\end{eqnarray}
The first term on the right hand side is $O(1/c)$ and the second scales like $\xi^{2(x(n)-\tilde x(n))}$. We therefore have the sum rule for the logarithmic correlation function
\be
\lambda(d-x)\int \langle D(r)D(0)\rangle_c d^dr
=-2x\langle D\rangle\log\xi+O(1)\,.
\ee

\section{Summary}\label{sec8}

In this article we explored the properties of logarithmic CFTs as limits of ordinary CFTs. In particular we showed how logarithmic behaviour arises in theories with an internal symmetry when the representation theory becomes singular. However these arguments also require some physical input as to which correlation functions should remain finite in the limit. 

We gave several examples of physical systems whose scaling limit is described by  CFTs exhibiting logarithmic behaviour in appropriate limits: quenched random magnets, the $O(n)$ model, and the $Q$-state Potts model. Although  we have focussed on the case when the central charge $c\to0$, these models also become logarithmic at other values of the parameters, depending on which observable is considered. 

Some of the physically measurable properties. such as the effective central charge, and the parameters $\beta$ and $\gamma$ in (\ref{323}, \ref{325}), for example, involve derivatives of parameters like $c$ and the scaling dimensions with respect to $n$ or $Q$. This casts doubt on the completeness of the physical description available by working directly within the logCFT, as is the case in supersymmetric approaches to disordered systems, for example.

We spent some time discussing how logCFTs with $c=0$ avoid the $c\to0$ paradox. The generic mechanism involves the collision of another operator $\widetilde T$ with the stress tensor $T$. In this way the operator $t\propto\widetilde T-T$ introduced by Gurarie \cite{Gur2} may be identified explicitly, and the parameter $b$ computed. However it should be noted that this $t$ transforms non-trivially under the internal symmetry group, which, for example, means that the non-logarithmic terms in its correlation functions are more complicated than those of a singlet operator. In addition, in two dimensions the bulk version of $t$ is not holomorphic. Thus many of the considerations of Ref.~\cite{GurLud} may apply only to boundary logCFTs.

We argued that the bulk $b$-parameter should be universal for a given logCFT, but that it may take different values in the bulk and on the boundary, as observed in two-dimensional models, and may even depend on the boundary condition.

Most of these ideas apply in arbitrary dimension $d$ (assuming that the scaling behaviour is described by a non-trivial RG fixed point.) 

Finally we considered massive logarithmic quantum field theories, and showed how the correlations of logarithmic operators differ from those of ordinary operators. We gave various adaptations of the $c$-theorem and $\Delta$-sum rules which show that universal properties of the ultraviolet logCFT, such as the $b$-parameter, may be extracted in principle from data in the massive theory.

It would be important to understand whether our approach extends to more interesting $c=0$ logCFTs, for example those based on sigma models over symmetric spaces \cite{sigma}, and how the group theoretic structure we have emphasised is reflected in the supersymmetric formulations of these problems.

\ack Some of the ideas here were revisited and further developed while I was participating in a workshop at the Institut Henri Poincar\'e, Paris, in 2011. I gratefully acknowledge its hospitality and the financial support of CNRS. Over the years I have had stimulating discussions about logCFT with Victor Gurarie, Ian Kogan, Hubert Saleur, Andreas Ludwig, Jesper Jacobsen and Romain Vasseur, among others. I thank David Ridout for useful comments on the first version of this paper.


\section*{References}

\begin{thebibliography}{1}
%
\bibitem{Weg} Wegner F J 1976 {\it Phase Transitions and Critical Phenomena} vol~6 ed C Domb and M S Green (New York: Academic Press) p~94
%
\bibitem{Sal} Rozansky L and Saleur H 1992 {\it Nucl. Phys. B} {\bf 376} 461; Saleur H 1992 {\it Nucl.Phys. B} {\bf 382} 486
%
\bibitem{Gur} Gurarie V 1993 {\it Nucl. Phys.} {\bf 410} 535
%
\bibitem{QH} Bhaseen M J \etal 2000 {\it Nucl. Phys. B} {\bf 580} 688;
Kogna I I and Tsvelik A M 2000 {\it Mod. Phys. Lett. A} {\bf 15} 931
%
\bibitem{Cardy1} Cardy J 1999 {\it Preprint} cond-mat/9911024
%
\bibitem{Cardy2} Cardy J 2001 {\it Statistical Field Theories (Proceedings of a NATO workshop, Como, June 2001)} ed A Cappelli and G Mussardo G ({\it Preprint} cond-mat/0111031)
%
\bibitem{Kogan} Kogan I I and Nichols A {\it Preprint} hep-th/0203207
%
\bibitem{orthog} Ferrara S Gatto A F and Grillo R 1972 {\it Nuov. Cim. A} {\bf 12} 959
%
\bibitem{FQS} Friedan D Qiu Z and Shenker S 1984 {\it Phys. Rev. Lett.} {\bf 52} 1575
%
\bibitem{Gur2} Gurarie V 1999 {\it Nucl. Phys. B} {\bf 546} 765
%
\bibitem{JCTT} Cardy J 1990 {\it Phys. Rev. Lett.} {\bf 65} 1443
%
\bibitem{Kom} Komargodski Z and Schwimmer A 2011 {\it JHEP} {\bf 1112} 099
%
\bibitem{JCcrossing} Cardy J 2002 {\it J. Phys. A} {\bf 35} L565
%
\bibitem{Ridout12} Ridout D 2012 {\it Preprint} arXiv:1203.3247
\bibitem{CG} Nienhuis B 1987 {\it Phase Transitions and Critical Phenomena} vol~11 ed C Domb and J L Lebowitz
(New York: Academic Press) p~1
%
\bibitem{DF} Dotsenko Vl S and Fateev V A 1984 {\it Nucl. Phys. B} {\bf 240} 312
%
\bibitem{JCsurf} Cardy J  1984 {\it Nucl. Phys. B} {\bf 240} 514
%
\bibitem{DupSal} Saleur H and Duplantier B 1987 {\it Phys. Rev. Lett.} {\bf 58} 2325;
Duplantier B and Saleur H 1987 {\it Nucl. Phys. B} {\bf 290} 291
%
\bibitem{GurLud} Gurarie V and Ludwig A W W 2002 {\it J. Phys. A} {\bf 35} L377
%
\bibitem{BCN} Bl\"ote H W J Cardy J L and Nightingale M P 1986 {\it Phys. Rev. Lett.} {\bf 56} 742
%
\bibitem{Aff} Affleck I 1986 {\it Phys. Rev. Lett.} {\bf 56} 746
%
\bibitem{JChigherd} Cardy J 1991 {\it Nucl. Phys. B} {\bf 366} 403
%
\bibitem{Harris} Harris A B 1974 {\it J. Phys. C} {\bf 7} 1761
%
\bibitem{Ludwig} Ludwig A W W 1987 {\it Nucl. Phys.}{\bf 285} 97
%
\bibitem{Brezin} Br\'ezin E and de Dominicis C 1998 {\it Europhys. Lett.} {\bf 44} 13
%
\bibitem{Davis} Davis T S and Cardy J 2000 {\it Nucl.Phys. B} {\bf 570} 713
%
\bibitem{EA} Edwards SF and Anderson P W 1975 {\it J. Phys. F} {\bf 5} 965
%
\bibitem{deGennes} de Gennes P G 1972 {\it Phys. Lett. A} {\bf 38} 339
%
\bibitem{V1} Vasseur R Jacobsen JL and Saleur H 2012 {\it J. Stat. Mech.} L07001
%
\bibitem{minus2} Balian R and Toulouse G 1973 {\it Phys. Rev. Lett.} {\bf 30} 544
%
\bibitem{V2} Vasseur R Gainutdinov AM Jacobsen JL and Saleur H 2012 {\it Phys. Rev. Lett.} {\bf 108} 161602
%
\bibitem{Doy} Doyon B Riva V and Cardy J 2006 {\it Comm. Math. Phys.} {\bf 268} 687
%
\bibitem{PottsRG} Amit D J 1976 {\it J. Phys. A } {\bf 9} 1441
%
\bibitem{cminus2} Ivashkevich E V 1999 {\it J . Phys. A} {\bf 32} 1691
%
\bibitem{SLE} See for example Cardy J 2005 {\it Ann. Phys.} {\bf 318} 81
%
\bibitem{SLEduality} Duplantier B 2000 {\it Phys. Rev. Lett.} {\bf 84} 1363
%
\bibitem{JCtorus} Cardy J 1986 {\it Nucl. Phys. B} {\bf 270} 186
%
\bibitem{Zam} Zamolodchikov AB 1986 {\it Pis'ma Zh. Eksp. Teor. Fiz.} {\bf 43} 565 ({\it JETP Lett.} {\bf 43} 730)
%
\bibitem{JCsumrule} Cardy J 1988 {\it Phys. Rev. Lett.} {\bf 60} 2709
%
\bibitem{Del} Cardy J Delfino G and Simonetti P 1996 {\it Phys. Lett. B} {\bf 387} 327
%
\bibitem{sigma} Zirnbauer M 1996 {\it J. Math. Phys.} {\bf 37} 4986; Altland A and Zirnbauer M 1997 
{\it Phys. Rev. B} {\bf 55} 1142
%
\end{thebibliography}
\end{document}